\documentclass[11pt]{article}
\usepackage{lineno}
\usepackage[a4paper, total={6in, 8in}]{geometry}
\usepackage[affil-it]{authblk}
\usepackage[english]{babel}
\usepackage{verbatim}
\usepackage{epsfig}
\usepackage{latexsym}
\usepackage{amsmath}
\usepackage{amssymb}
\usepackage{amsfonts}
\usepackage{graphicx}
\usepackage{appendix}
\usepackage[dvipsnames]{xcolor}
\usepackage{multirow}
\usepackage{blindtext, rotating}
\usepackage[pdftex]{pict2e}
\usepackage{slashed}
\usepackage{physics}
\usepackage{leftidx}

\numberwithin{equation}{section}

\usepackage{url}

\usepackage[colorlinks=true]{hyperref}
\hypersetup{
linkcolor=BrickRed
,citecolor=Green
,filecolor=Mulberry
,urlcolor=NavyBlue
,menucolor=BrickRed
,runcolor=Mulberry
,linkbordercolor=BrickRed
,citebordercolor=Green
,filebordercolor=Mulberry
,urlbordercolor=NavyBlue
,menubordercolor=BrickRed
,runbordercolor=Mulberry
}

\usepackage{cleveref}

\usepackage{csquotes}
\usepackage[
backend=bibtex,
style=numeric,
doi=true,
sorting=none,
url=false
]{biblatex}

\usepackage{comment}

\newcommand{\lb}{\left\lbrace}
\newcommand{\rb}{\right\rbrace}
\newcommand{\bstheta}{\boldsymbol{\theta}}
\newcommand{\bsy}{\boldsymbol{y}}

\title{\bfseries A regularization-independent Information Criterion for correlated data}
\author{Davide Laudicina%
    \thanks{Email address: \href{mailto:davide.laudicina@ruhr-uni-bochum.de}{davide.laudicina@ruhr-uni-bochum.de}}}
\affil{\small Fakultät für Physik und Astronomie, Institut für Theoretische Physik II,\\Ruhr-Universität Bochum, 44780 Bochum, Germany}
\date{}

\begin{document}
\maketitle

\begin{center}
\rule{\textwidth}{0.4pt}    
\end{center}

\begin{abstract}
We generalize the Akaike Information Criterion (AIC) to allow model and data subset selection
when regularized fits are performed on correlated data. 
While the standard generalization to regularized fits is inherently dependent on
the type of regularization applied to the data covariance matrix, our definition, the RIAIC,
is, on average, regularization-independent. The regularization-independence is achieved
by defining the new Information Criterion as an estimator of the model-dependent contribution
to the correlated Kullback--Leibler divergence. The Bayesian extension is derived as well.
We test the new criterion on lattice QCD data by performing regularized fits
and show that it reproduces the selection properties of the standard correlated AIC,
with fluctuations which are compatible with the expected statistical uncertainties.
These results enable the application of subset and model selection techniques to cases
where regularizations are applied to the covariance matrix of the fitted dataset.
\end{abstract}

\begin{center}
\rule{\textwidth}{0.4pt}    
\end{center}

\newpage
\tableofcontents

\flushbottom

\section{Introduction}
\label{sec:introduction}
The goal of identifying the optimal description of a dataset 
plays a pivotal role in both experimental and theoretical data
analysis. By \emph{optimal} we refer to a parameterization that simultaneously maximizes the information extracted from the data 
while minimizing the complexity of the model itself. In addition to 
the optimal parameterization for a fixed dataset, comparisons of different data subsets are
also necessary. More generally, model 
and data subset selection provides a valuable tool to assess and 
quantify systematic uncertainties arising from arbitrary choices on the model
and on the fitted dataset.

Information-Theoretic Criteria (ICs) are a widely used approach to model selection problems. They are theoretically sound, since their definition is rooted in information theory through the Kullback--Leibler divergence, and, by construction, they represent a trade-off between quality of the fit and complexity of the description.

One commonly used IC is the Akaike Information Criterion (AIC)~\cite{Akaike1,Akaike2}.
The standard definition of the AIC explicitly assumes that correlations in the dataset under study are properly taken into account.

However, it is often the case that covariance matrices
are estimated on a limited, and noisy, sample of data.
This may lead to close-to-singular covariance matrices,
for which minimization procedures, such as least squares methods,
become impractical. 
In this context, a possible solution to 
extract additional information from the data is to perform fits 
by employing regularized versions of the covariance matrix.
Typical regularizations include the
application of SVD cuts or the limiting case of a fully uncorrelated
fit with a diagonal covariance matrix.

While a proper treatment of the covariance matrix
and the impact of regularizations can be addressed at the level
of the quality of the fit, by redefining the expectation values 
of chi-square variables, see Ref.~\cite{Bruno:2022mfy},
accounting for these effects at the level of the AIC
still represents a gap which prevents a reliable comparison
of models and subsets in the case of regularized fits.
In general, ICs are rooted in information theory and regularizations of the
covariance matrix effectively modify the amount of information
contained in the selected dataset, by removing poorly determined information.
Similarly, in a Bayesian framework, the use of priors
modifies the information used as input to the fit, by
adding knowledge on the system under study.
On the basis of these considerations, the application of any
regularization to the data covariance matrix introduces
a dependence on the regularization itself in the AIC. As such,
a systematic comparison of models and datasets when fits are performed with regularized
covariance matrices is still not possible.
This is particularly relevant when the size of the fitted dataset is varied, since the 
regularization required to obtain a stable fit typically changes with it.
A typical example in the analysis
of lattice field theory data is represented by the search for the optimal fit window of
a correlation function, where different windows may require different regularizations, mostly
depending on the size of the window itself.
Using the approach proposed here, and detailed in the following sections,
the regularization of the covariance matrix, which is necessary to stabilize the fit,
is incorporated into the AIC.

This work is organized as follows. In Sec.~\ref{sec:AIC} we briefly review the idea behind Information Criteria, by focusing on the definition of the Akaike Information Criterion and its extension to dataset selection. In the same section, we report the standard generalization of the AIC to 
regularized fits and we show that its extension to subset selection problems induces an explicit dependence on the regularization which makes it unreliable.
In Sec.~\ref{sec:RIAIC} we propose a new Information Criterion which is explicitly built to be, on average, regularization-independent.
Finally, in Sec.~\ref{sec:example} we explicitly carry out numerical tests on actual lattice data to validate the formulae derived in this work.

The main results of this work are represented by the expressions in eqs.~\eqref{eq:gAIC} and~\eqref{eq:gAIC_full},
together with their Bayesian extension in eq.~\eqref{eq:RIBAIC2}.

Throughout this work we often refer to examples or typical situations encountered in lattice field theory studies, e.g. the identification of the optimal fit window in the 
extraction of the energy spectrum from Euclidean lattice correlators. Nevertheless the formulae derived here apply to any situation 
in which the data follow a multivariate Gaussian distribution.

\section{Information-Theoretic Criteria}
\label{sec:AIC}
Given a dataset $\bsy=\lb y_1, \dots, y_{n_d} \rb$, and a model function $\eta$ which depends on the parameter vector $\bstheta=\lb \theta_1,\dots,\theta_{n_\eta}\rb$ and, optionally, on some other variables, let us consider a set of candidate model probabilities, i.e. a set of probability density functions~\cite{https://doi.org/10.1002/wics.1460}
\begin{align}
{\cal M}_{\Sigma} \, = \, \lb {\cal L}(\bsy, \Sigma \mid \bstheta, \eta) \bigm| \bstheta \in \Theta_\eta \rb
\end{align}
where $\Theta_\eta$ is the parameter space for a given model function $\eta$ and ${\rm dim}\left[\Theta_\eta\right] = n_\eta$. The \emph{true} covariance matrix $\Sigma$ is 
in most cases unknown and it is usually replaced by the sample covariance matrix~\cite{bootstrap,Wolff:2003sm} which provides a consistent and asymptotically unbiased estimator of $\Sigma$.

Given a candidate probability density function ${\cal L} \in{\cal M}_{\Sigma}$ and the data-generating unknown distribution of the dataset $P(\bsy)$, we define the corresponding Kullback--Leibler (KL) divergence~\cite{10.1214/aoms/1177729694} as
\begin{align}
\label{eq:KL}
D_{\rm KL} (P\mid\mid{\cal L}) \, &= \, \int d\bsy P(\bsy) \log \left[ \frac{P(\bsy)}{{\cal L}(\bsy, \Sigma \mid \bstheta, \eta)} \right] \, \\[.2cm] 
& = \, \mathbb{E} \big[ \log P(\bsy) \big] + \mathbb{E} \big[-\log {\cal L}(\bsy, \Sigma \mid \bstheta,\eta) \big] \,, \nonumber
\end{align}
where in the last line $\mathbb{E}[\cdot]$ denotes the expectation value over the true probability distribution $P(\bsy)$. Note that $D_{\rm KL} (P\mid\mid{\cal L})=0$ if and only if ${\cal L} = P$. Therefore the empirical description that best approximates the true probability distribution is found by minimizing the KL divergence, and eq.~\eqref{eq:KL} can be interpreted as a measure of how large the deviations between the candidate and the true distribution are.

The first term appearing in eq.~\eqref{eq:KL} is, by definition, independent of the candidate model. Therefore, minimizing the Kullback--Leibler divergence with respect to the parameter vector $\bstheta$ is equivalent to minimizing the model-dependent quantity
\begin{align}
\label{eq:Delta}
\Delta(\Sigma \mid \bstheta,\eta) \, = \, \mathbb{E} \big[ -2\log {\cal L}(\bsy, \Sigma \mid \bstheta,\eta)  \big] \, ,
\end{align}
where the additional factor of 2 is conventional.
Let us stress that the expression above depends on the data-generating distribution $P(\bsy)$ and it is, therefore, not directly accessible. However, given an estimator $\bstheta^\star$ for the parameter vector $\bstheta$, Information Criteria are defined as unbiased \emph{plug-in}~\cite{Neil:2022joj,Neil:2023pgt} estimators of eq.~\eqref{eq:Delta} as
\begin{equation}
\label{eq:IC}
{\rm IC}_\eta \, = \, -2\log {\cal L}(\bsy,\Sigma \mid \bstheta^\star,\eta) + {\cal B}(\bsy, \Sigma,\eta) \, ,
\end{equation}
where ${\cal B}(\bsy,\Sigma,\eta)$ is a bias term that depends on the dataset and on the parameter space associated with the candidate model ${\cal L}(\bsy,\Sigma \mid \bstheta,\eta)$ and 
its expression depends on the choice of the plug-in estimator $\bstheta^\star$ used in eq.~\eqref{eq:IC}. In eq.~\eqref{eq:IC} the subscript indicates the model function. As usual, the bias is computed as the expectation value of the difference between the biased estimator and the target quantity, i.e.
\begin{align}
\label{eq:bias}
{\cal B}(\bsy, \Sigma,\eta) \equiv \mathbb{E} \Big[ 2\log {\cal L}(\bsy,\Sigma \mid \bstheta^\star,\eta) + \Delta(\Sigma \mid\bstheta^\star,\eta) \Big] \, .
\end{align}
As we will see in the following, the bias in eq.~\eqref{eq:bias} usually plays the role of a penalty term which penalizes overly complex models and favors simpler parameterizations for a given dataset; in this sense, the bias naturally discourages overfitting.

\subsection{Akaike Information Criterion}
In this section we briefly review the standard definition of the Akaike Information Criterion (AIC)~\cite{Akaike1,Akaike2}. For a detailed introduction to the topic we refer to Ref.~\cite{https://doi.org/10.1002/wics.1460}.
In the following we work under the assumption that the model is correctly or over-specified.
In practice, a model can be considered correctly specified if the corresponding fit is accepted based
on its quality, e.g. if the corresponding $p$-value is above a given threshold. For regularized fits the quality of the fit must be computed as in Ref.~\cite{Bruno:2022mfy}.
In this case the data-generating distribution belongs to the set ${\cal M}_{\Sigma}$ and can be written as $P(\bsy) = {\cal L} \left( \bsy, \Sigma \mid \bstheta_0, \eta \right)$, where $\bstheta_0$ is the true value of the parameter vector.

The standard definition of the AIC assumes as plug-in estimator for the parameter vector $\bstheta$ the maximum likelihood estimator (MLE) $\bstheta_{\rm MLE}^\text{\tiny{$\Sigma$}} \equiv {\rm argmax}_{\bstheta}\Big[{\cal L}(\bsy, \Sigma \mid \bstheta,\eta)\Big]$ which is asymptotically unbiased.
Following the definitions given in the previous section, the negative log-likelihood evaluated at the MLE parameter defines an Information Criterion according to eq.~\eqref{eq:IC} with bias computed as in eq.~\eqref{eq:bias}. 
In Refs.~\cite{Akaike1,Akaike2} it was shown that the bias term has a particularly simple expression and depends only on the overall dimension of the parameter space $\Theta_\eta$, leading to
\begin{align}
\label{eq:AIC_bias}
{\cal B}(\bsy, \Sigma,\eta) \equiv \mathbb{E} \Big[ 2\log {\cal L}(\bsy,\Sigma \mid \bstheta^\text{\tiny{$\Sigma$}}_{\rm MLE},\eta) + \Delta(\Sigma \mid\bstheta^\text{\tiny{$\Sigma$}}_{\rm MLE},\eta) \Big] = 2 n_\eta \, .
\end{align}
For a detailed derivation of the bias see Ref.~\cite{https://doi.org/10.1002/wics.1460}. Therefore an unbiased estimator for eq.~\eqref{eq:Delta} is given by the so-called Akaike Information Criterion \cite{Akaike1,Akaike2}, which is defined as 
\begin{equation}
\label{eq:AIC}
{\rm AIC}_\eta = -2\log{\cal L}(\bsy,\Sigma \mid \bstheta_{\rm MLE}^\text{\tiny{$\Sigma$}},\eta) + 2n_\eta \,.
\end{equation}
Under the assumption that both the data-generating and the candidate model probabilities are multivariate Gaussian distributions, the AIC can be written, up to terms which are constant across all models, in the standard way as
\begin{align}
{\rm AIC}_\eta = \chi^2_{\text{\tiny{$\Sigma$}}} + 2n_\eta \, ,
\end{align}
where $\chi^2_{\text{\tiny{$\Sigma$}}}$ is the chi-square evaluated at the MLE parameters and the subscript makes explicit the
covariance matrix used in its definition.

Therefore, smaller values of the AIC correspond to smaller values of eq.~\eqref{eq:Delta} and, as a consequence, a better agreement between the candidate model and the true probability distribution. Note that, in general, it holds $\mathbb{E}\left[ {\rm AIC}_\eta \right] = n_d + n_\eta$. Therefore parameterizations with a large number of parameters will be penalized, if the quality of the fit is similar. In this sense, the AIC is a trade-off between the quality of the fit and the complexity of the model.

\subsubsection{Model and dataset selection}
\label{sec:dataset_selection}
The problem of data subset selection, i.e. the problem of finding the most informative data subset which can be described by the simplest model, can be rephrased in terms of model selection problems as detailed in Refs.~\cite{Neil:2022joj,Neil:2023pgt}.

The core idea is that a given dataset, which hereafter we refer to as the \emph{full dataset}, can be split into two different subsets: one which is included in the actual fit and one which is excluded. Under the assumption that the fitted and non-fitted datasets are statistically independent, the total KL divergence and, as a consequence, eq.~\eqref{eq:Delta} factorize into a sum of two independent terms
\footnote{The extension to correlated fitted and non-fitted subsets is left for a future work; see Refs.~\cite{Jay:2020jkz,Neil:2022joj} for a further discussion.}.
This leads to a factorization of the corresponding AIC as
\begin{align}
{\rm AIC}_{{\rm full}} = {\rm AIC}_{\eta} + {\rm AIC}_{\text{\tiny C}}
\end{align}
where the first term on the r.h.s. corresponds to the AIC of the fitted dataset $\bsy$ where the model function $\eta$ has been used to describe the data and ${\rm AIC}_{\text{\tiny C}}$ is the AIC of the non-fitted dataset\footnote{Here we partially depart from the standard nomenclature. In the literature the AIC of the fitted dataset is often called ${\rm AIC}_{\text{\tiny K}}$ where the label ``K" refers to the \emph{kept} dataset, while ${\rm AIC}_{\text{\tiny C}}$ refers to the \emph{cut} dataset, i.e. the one excluded from the actual fit.}.

In the literature different definitions for ${\rm AIC}_{\text{\tiny C}}$ have been proposed, see Refs.~\cite{Borsanyi:2020mff,Jay:2020jkz} and Ref.~\cite{Neil:2023pgt} for a detailed comparison. The results presented here are independent of the chosen definition; nevertheless, in the following we will always assume that the \emph{perfect model} definition is employed~\cite{Jay:2020jkz}. The generalization to other definitions is straightforward.

The perfect model is a model for the non-fitted dataset with as many parameters as data points, e.g. a polynomial
interpolant. Therefore, under the Gaussian approximation of the non-fitted dataset, since the perfect model provides
a vanishing chi-square, the AIC associated with the non-fitted subset reduces to
\begin{align}
\label{eq:pm}
{\rm AIC}_{\text{\tiny C}} = 2n_{\text{\tiny C}} + \log \det \Sigma_{\text{\tiny C}} + n_{\text{\tiny C}} \log 2\pi \, ,
\end{align}
where the first term on the r.h.s. is twice the number of parameters of the perfect model, which is, by definition, equal to the number of data points of the non-fitted dataset and, most importantly, it does not depend on the explicit functional form of the perfect model. As a consequence, for normally-distributed data the AIC of the full dataset can be written as
\begin{align}
\label{eq:AIC_full2}
{\rm AIC}_{\rm full} = \chi^2_{\text{\tiny{$\Sigma$}}} + 2n_\eta + 2n_{\text{\tiny C}} + \log \det \Sigma_{\rm full} + n_{\rm full} \log 2\pi, 
\end{align}
where $n_{\rm full} = n_{d} + n_{\text{\tiny C}}$ and we employed the fact that the fitted and non-fitted data subsets are statistically uncorrelated, i.e. $\Sigma_{\rm full} = {\rm diag}\left(\Sigma, \Sigma_{\text{\tiny C}} \right)$. Finally, by neglecting terms which are common to all models and to the full dataset, it simplifies to the known result \cite{Neil:2023pgt,Jay:2020jkz}
\begin{align}
\label{eq:AIC_full}
{\rm AIC}_{\rm full} = \chi^2_{\text{\tiny{$\Sigma$}}} + 2n_\eta + 2n_{\text{\tiny C}} \, .
\end{align} 
For a specific full dataset and model function $\eta$, eq.~\eqref{eq:AIC_full} can be used to extend model selection problems to subset selection ones. Specifically, the use of ${\rm AIC}_{\rm full}$ will automatically promote models which provide a good description of a data subset which is as large as possible, with a model depending on a number of parameters which is as small as possible.

\subsection{Akaike Information Criterion and regularized fits}
\label{sec:UAIC}
The definition of the AIC given in eq.~\eqref{eq:AIC} can be easily extended to the scenario where fits with regularized covariance matrices are performed on correlated, normally distributed, data. This result is reported in Ref.~\cite{Frison:2023lwb} and in this section we briefly review the framework in which it is derived.

As before, we assume that the model $\eta$ is correctly specified and the true probability distribution is a multivariate normal distribution $P(\bsy) = {\cal L} \left( \bsy, \Sigma \mid \bstheta_0, \eta \right)$, where $\bstheta_0$ is the true value of the parameter vector.
If the covariance matrix $\Sigma$ is numerically ill-conditioned, it is common practice to perform the fit using 
a regularized matrix.
Therefore, the plug-in estimator used to estimate eq.~\eqref{eq:Delta} is $\bstheta_{\rm MLE}^\text{\tiny{$C$}} \equiv {\rm argmax}_{\bstheta}\Big[{\cal L}(\bsy, C \mid \bstheta,\eta)\Big]$
which is, unlike $\bstheta_{\rm MLE}^\text{\tiny{$\Sigma$}}$, obtained by maximizing the log-likelihood using a generic positive-definite matrix $C$ whose elements are $O(N^{-1})$, where $N$ is the size of the Monte Carlo sample. Note that, in general, these two estimators might differ and their difference depends on how much the
regularized covariance matrix $C$ differs from $\Sigma$.

In this context, we are testing model probabilities which are no longer in the set ${\cal M}_{\Sigma}$ but in the much larger set ${\cal M}_{C}$ which contains ${\cal M}_{\Sigma}$ as the specific case of models based on correlated fits.
By defining the estimator of the regularized model-dependent term of the KL-divergence as in eq.~\eqref{eq:Delta}, the bias to be computed is
\begin{align}
\label{eq:bias_UAIC}
{\cal B}(\bsy, C,\eta) \equiv \mathbb{E} \Big[ 2\log {\cal L}(\bsy,C \mid \bstheta_{\rm MLE}^\text{\tiny{$C$}},\eta) + \Delta(C \mid\bstheta_{\rm MLE}^\text{\tiny{$C$}},\eta) \Big] = 2n_\eta^{\rm eff} \, ,
\end{align}
where, in analogy with the standard expression in eq.~\eqref{eq:AIC}, we introduced the effective number of parameters as
\begin{align}
\label{eq:neff}
n_\eta^{\rm eff} = \tr \left[ C^{-1/2} \Sigma C^{-1/2} {\cal P}\right] \, ,
\end{align}
where the projector ${\cal P}$ is defined in Ref.~\cite{Bruno:2022mfy} and satisfies $\tr[{\cal P}] = n_\eta$, see eq.~\eqref{eq:proj}.
By using this expression for the bias, the regularized version of the AIC, which hereafter we refer to as $\overline{\operatorname{AIC}}$ in order to distinguish it from the standard definition, reads
\begin{align}
\label{eq:UAIC1}
\overline{\operatorname{AIC}}_\eta = -2\log{\cal L}(\bsy,C\mid \bstheta_{\rm MLE}^\text{\tiny{$C$}},\eta) + 2 n_\eta^{\rm eff} \, .
\end{align}
The expression in eq.~\eqref{eq:UAIC1} corresponds, up to constant terms, to the one reported in Ref.~\cite{Frison:2023lwb}. Notice that the definition of the $\overline{\operatorname{AIC}}$ is inherently regularization-dependent and, in particular, under the assumption of normally-distributed data, its expectation value reads
\begin{align}
\label{eq:UAIC_exp}
    {\mathbb{E}} \left[ \overline{\operatorname{AIC}}_\eta \right] = n_d^{\rm eff} + n_{\eta}^{\rm eff} + \log\det C + n_d\log 2\pi \, ,
\end{align}
where $n_d^{\rm eff} = \tr\left[ C^{-1} \Sigma \right]$.
Therefore, by applying different regularizations to the covariance matrix, the $\overline{\operatorname{AIC}}$ will yield different values. This is not surprising, since eq.~\eqref{eq:UAIC1} is an estimate of how much the regularized candidate model and the true model probability differ. 
As a final remark, note that in the correlated limit, i.e. for $C = \Sigma$, the expression in eq.~\eqref{eq:UAIC1} reproduces exactly the standard result in eq.~\eqref{eq:AIC}.

\subsubsection{Dataset selection and regularized fits}
\label{sec:UAIC_dataset}
In this section we show that the definition given in eq.~\eqref{eq:UAIC1}, which naturally arises when regularized fits are performed on correlated data, is incompatible with the type of dataset selection problems described in Sec.~\ref{sec:dataset_selection}. The typical case is encountered when some sort of regularization is applied to the covariance matrix of a fitted subset $\Sigma$, while the covariance matrix of the non-fitted dataset is left untouched. In this case, the expression for ${\rm AIC}_{\text{\tiny C}}$ is not altered with respect to eq.~\eqref{eq:pm}. This leads to
\begin{align}
\label{eq:uAIC_full2}
\overline{\operatorname{AIC}}_{\rm full} = \chi^2_{C} + 2n_{\eta}^{\rm eff} + 2n_{\text{\tiny C}} + \log\det C_{\rm full} + n_{\rm full} \log 2\pi \, ,
\end{align}
where $C_{\rm full} = {\rm diag}\left( C,\Sigma_{\text{\tiny C}} \right)$. In general the term $\log\det C_{\rm full}$ is no longer common to all the subsets of the full dataset, since $C$ explicitly depends on the regularization employed on the chosen fitted subset. Therefore, in model and data subset selection procedures which involve regularized fits, by removing terms which are common to the full dataset, the estimator to be used is
\begin{align}
\label{eq:uAIC_full}
\overline{\operatorname{AIC}}_{\rm full} = \chi^2_{C} + 2n_{\eta}^{\rm eff} + 2n_{\text{\tiny C}} + \log\det C_{\rm full} \, ,
\end{align}
in contrast to the standard case of correlated fits, where the simpler formula in eq.~\eqref{eq:AIC_full} applies. 

As mentioned in Sec.~\ref{sec:introduction}, regularizing the covariance matrix induces a loss of information.
In the expression above part of this loss of information is encoded in the different determinant appearing in the expression of the $\overline{\operatorname{AIC}}$ and can be easily understood if the applied regularization involves an SVD cut
on the smallest eigenvalues of $\Sigma$ in order to make it less ill-conditioned. In this case, it follows that 
$\det \Sigma_{\rm full} < \det C_{\rm full}$ and the definition in eq.~\eqref{eq:uAIC_full}
always promotes models and datasets with a regularized covariance matrix $C$ closer to $\Sigma$.
The crucial point is that if the null-hypothesis is not rejected, i.e. there is no statistical evidence that
the model is misspecified,
then $\bstheta_{\rm MLE}^\text{\tiny{$C$}}$ is still a good estimator of the true parameter vector $\bstheta_0$.
This implies that the information lost in the regularization procedure only plays the role of stabilizing the fit
and it is not needed to have a good estimator of the parameter vector.
The main reason is that the removed information is poorly estimated since the smallest eigenvalues
of the covariance matrix are usually affected by large errors.

Therefore, by construction, the $\overline{\operatorname{AIC}}$ in eq.~\eqref{eq:uAIC_full} does not allow one to compare subsets fitted
with different regularizations and the standard trade-off between complexity and quality of the fit is no longer the main feature
of the $\overline{\operatorname{AIC}}$. The regularization dependence of eqs.~\eqref{eq:UAIC1} and~\eqref{eq:uAIC_full} and its effects are illustrated in Sec.~\ref{sec:regularization} on a fixed dataset and in Secs.~\ref{sec:dataset} and~\ref{sec:dataset_and_model} in the context of dataset and model selection, respectively.

\section{Regularization-independent AIC}
\label{sec:RIAIC}
We tackle the problem outlined in the previous section by deriving an expression of the Akaike Information Criterion which is, on average, independent of the chosen regularization.
This is obtained by defining an estimator of eq.~\eqref{eq:Delta} which is able to account for the loss of information due to the regularization of the covariance matrix.

The regularization dependence of the $\overline{\operatorname{AIC}}$ arises from the fact that the expression in eq.~\eqref{eq:UAIC1} 
has been derived as the unbiased estimator of the regularized version of eq.~\eqref{eq:Delta}, i.e. computed with covariance matrix $C$ instead of the standard $\Sigma$. To cope with this issue, in the following we define a regularization-independent Information Criterion as an estimator of the \emph{correlated} $\Delta(\Sigma \mid \bstheta,\eta)$. This allows us to absorb the loss of information due to the regularization applied to the covariance matrix into the bias and, therefore, to get an expression which is, on average, regularization-independent.
We now take the plug-in estimator $\bstheta_{\rm MLE}^\text{\tiny{$C$}}$; therefore the Information Criterion reads
\begin{align}
\label{eq:RIAIC1}
    {\rm RIAIC}_\eta =  -2\log{\cal L}\left( \bsy,C \mid \bstheta_{\rm MLE}^\text{\tiny{$C$}},\eta \right) + {\cal B}\left(\bsy,C,\eta \right) \, ,
\end{align}
where ${\rm RIAIC}$ stands for \emph{regularization-independent} AIC.
The crucial step to absorb any regularization-dependent term is taking the regularized log-likelihood in the equation above as an estimator of eq.~\eqref{eq:Delta}. The bias to be computed reads
\begin{align}
\label{eq:bias_gAIC}
{\cal B}(\bsy, C,\eta) \equiv \mathbb{E} \Big[ 2\log {\cal L}(\bsy,C \mid \bstheta_{\rm MLE}^\text{\tiny{$C$}},\eta) + \Delta(\Sigma \mid\bstheta_{\rm MLE}^\text{\tiny{$\Sigma$}},\eta) \Big] \, .
\end{align}
Note that the second term on the r.h.s. is exactly the term appearing in the bias of the standard AIC in eq.~\eqref{eq:AIC_bias} and is evaluated at the correlated MLE estimator, i.e. $\bstheta_{\rm MLE}^\text{\tiny{$\Sigma$}}$. This is in contrast to eq.~\eqref{eq:bias_UAIC} where it is evaluated at the regularized MLE.

If the model is correctly specified, under the assumption of normally-distributed data, see App.~\ref{sec:AppA} for the detailed derivation, the bias reads
\begin{align}
\label{eq:bias_gAIC_2}
{\cal B}(\bsy, C,\eta) =  n_d  - n_d^{\rm eff} + n_\eta + n_\eta^{\rm eff} + \log\det \Sigma -\log\det C \,,
\end{align}
where $n_\eta^{\rm eff}$ and $n_d^{\rm eff}$ have been introduced in eqs.~\eqref{eq:neff} and~\eqref{eq:UAIC_exp}. This leads to the expression of the regularization-independent AIC
\begin{align}
\label{eq:gAIC}
{\rm RIAIC}_\eta &= \chi^2_C + n_d - n_d^{\rm eff} + n_\eta + n_\eta^{\rm eff} + n_d \log 2\pi + \log\det \Sigma \,.
\end{align}
Note that the regularization-dependent term $\log\det C$ appearing in the bias in eq.~\eqref{eq:bias_gAIC_2} exactly cancels out the contribution arising from the log-likelihood in eq.~\eqref{eq:RIAIC1}.
Similarly to the $\overline{\operatorname{AIC}}$ in eq.~\eqref{eq:UAIC1}, the formulation in eq.~\eqref{eq:gAIC} reproduces, in the correlated limit $C=\Sigma$, the standard AIC in eq.~\eqref{eq:AIC}.
However, unlike the $\overline{\operatorname{AIC}}$, see eq.~\eqref{eq:UAIC_exp}, the expectation value of eq.~\eqref{eq:gAIC} is exactly the one of eq.~\eqref{eq:AIC}, i.e.
\begin{align}
    {\mathbb{E}} \left[ {\rm RIAIC}_\eta \right] = n_d + n_\eta + \log\det\Sigma + n_d\log2\pi \,,
\end{align}
and it only depends on the non-regularized dataset and on the number of parameters of the model. 
In this sense, eq.~\eqref{eq:gAIC} is, on average, regularization-independent and deviations from eq.~\eqref{eq:AIC} might arise only due to fluctuations in the quality of the fit.
The size of the statistical fluctuations can be quantified by computing the variance, for a given fitted model,
of $\Delta_{\text{\tiny{RIAIC}}}={\rm RIAIC}-{\rm AIC}$; see eq.~\eqref{eq:variance}
for the full expression and App.~\ref{sec:AppC} for the derivation.

The idea behind eq.~\eqref{eq:gAIC} is that, provided that the null-hypothesis has not been rejected, the maximum likelihood estimator $\bstheta_{\rm MLE}^\text{\tiny{$C$}}$ is still a good estimator of the true parameter vector $\bstheta_0$. Therefore, the regularization of the covariance matrix $\Sigma$ is only relevant for the convergence of the fit but the poorly-estimated information which is lost in the regularization step is irrelevant for the purpose of obtaining a reliable estimator of $\bstheta_0$.
In practice, care must be taken in choosing the regularization of the covariance matrix. Indeed, this argument relies on the fact that the regularization
only removes poorly-estimated information; however, if the regularization removes information which is relevant for a correct estimation of the model
parameters, the RIAIC may deviate from the corresponding correlated AIC.

The criterion in eq.~\eqref{eq:gAIC}, being on average regularization-independent, allows the comparison between models fitted on data subsets which are regularized in different ways; see Sec.~\ref{sec:example} for some detailed examples.

\subsection{Extension to dataset selection}
As shown in Sec.~\ref{sec:UAIC_dataset}, once we try to rephrase data subset selection problems in terms of model selection ones, the definition of the $\overline{\operatorname{AIC}}$ fails to provide the usual trade-off criterion between quality of the fit and complexity; see Sec.~\ref{sec:dataset} for a detailed example. This issue is solved if the definition in eq.~\eqref{eq:gAIC} is used instead. As before, we assume that the covariance matrix of the non-fitted dataset $\Sigma_{\text{\tiny C}}$ is not regularized. Under this assumption, the bias correction in eq.~\eqref{eq:bias_gAIC_2} absorbs the loss of information due to the regularization of the covariance matrix of the fitted subset and the ${\rm RIAIC}_{\rm full}$ depends only on the determinant of the original covariance matrix $\Sigma_{\rm full} = {\rm diag} \left( \Sigma,\Sigma_{\text{\tiny C}}\right)$, i.e.
\begin{align}
\label{eq:gAIC_full2}
{\rm RIAIC}_{\rm full} &= \chi^2_C + n_d - n_d^{\rm eff}  + n_\eta + n_\eta^{\rm eff} + 2n_{\text{\tiny C}} + \log\det \Sigma_{\rm full} + n_{\rm full} \log 2\pi ,
\end{align}
where, again, we used the perfect-model prescription for the AIC of the non-fitted dataset, but generalizations to other definitions are possible as well.
Therefore, by removing terms which are common to the full dataset and all models we obtain the final expression
\begin{align}
\label{eq:gAIC_full}
{\rm RIAIC}_{\rm full} = \chi_C^2 + n_d - n_d^{\rm eff} + n_\eta + n_\eta^{\rm eff}+2n_{\text{\tiny C}} \, ,
\end{align}
which does not depend on the log-determinant of the regularized covariance matrix, unlike eq.~\eqref{eq:uAIC_full}.
This expression inherits all the properties of the RIAIC: it reproduces, in the correlated limit, the standard result in eq.~\eqref{eq:AIC_full} and its expectation value is regularization-independent.
\subsection{Extension to the Bayesian framework}
\label{sec:bayes_extension}
Assume that the \emph{a priori} knowledge about some of the parameters of the model function $\eta$ is encoded in $n_\pi$ Gaussian priors, with $n_\pi\leq n_\eta$, characterized by mean $\tilde\bstheta$ and covariance matrix $C_\theta$. Then the posterior probability distribution, up to an overall normalization factor, reads
\begin{align}
\label{eq:bayes_L}
p \left( \bstheta \mid \bsy\right) = {\cal L}\left( \bsy,C \mid \bstheta,\eta \right) \times  \pi \left( \bstheta; \tilde\bstheta, C_\theta \right) \, .
\end{align}
In Refs.~\cite{Neil:2022joj,Jay:2020jkz} different possible extensions of the AIC to the Bayesian framework are presented. Given the inherently frequentist nature of the AIC, here we generalize to our framework what the authors of Ref.~\cite{Neil:2022joj} call BAIC, namely the \emph{Bayesian} AIC; however, the extension to other proposals is straightforward.
We use as plug-in estimator for the Information Criterion the maximum a posteriori (MAP) estimator obtained from a regularized fit, i.e. $\bstheta_{\rm MAP}^\text{\tiny{$C$}} = {\rm argmax}_{\bstheta} \lb {p} \left( \bstheta\mid \bsy\right) \rb$. In this way the regularization-independent BAIC (RIBAIC) reads
\begin{align}
\label{eq:RIBAIC}
    {\rm RIBAIC}_\eta =  -2\log{\cal L}\left( \bsy,C \mid \bstheta_{\rm MAP}^\text{\tiny{$C$}},\eta \right) + {\cal B}\left(\bsy,C,\eta \right) \, .
\end{align}
Similarly to the frequentist case in eq.~\eqref{eq:bias_gAIC}, any regularization-dependent term is removed by the bias
\begin{align}
\label{eq:RIBAIC_bias}
{\cal B}(\bsy, C,\eta) \equiv \mathbb{E} \Big[2\log {\cal L}(\bsy,C \mid \bstheta_{\rm MAP}^\text{\tiny{$C$}},\eta) + \Delta(\Sigma \mid\bstheta_{\rm MLE}^\text{\tiny{$\Sigma$}},\eta) \Big]  \, .
\end{align}
It is worth noticing that, in contrast to the frequentist case, the estimator $\bstheta_{\rm MAP}^\text{\tiny{$C$}}$ does not minimize the chi-square appearing in the log-likelihood in eq.~\eqref{eq:RIBAIC}, but rather the so-called \emph{augmented} chi-square \cite{Lepage:2001ym}, which encompasses both the data and the prior contributions. Nevertheless, if the model, including the priors, is correctly specified, the difference between $\bstheta_{\rm MAP}^\text{\tiny{$C$}} $ and $\bstheta_{\rm MLE}^\text{\tiny{$C$}}$ is expected to be small, leading to a bias correction which is numerically close to the one in eq.~\eqref{eq:bias_gAIC_2}. The derivation of the full bias is carried out in App.~\ref{sec:AppB} and leads, for normally-distributed data, to the expression
\begin{align}
\label{eq:RIBAIC2}
    {\rm RIBAIC}_\eta = \chi_C^2 + n_d - n_d^{\rm eff} + n_\eta + \hat{n}_\eta^{\rm eff} + n_d\log 2\pi +  \log\det\Sigma - \varepsilon({\tilde\bstheta}) \, ,
\end{align}
where we introduced the effective number of parameters
\begin{align}
    \hat{n}_{\eta}^{\rm eff} = 2\tr\left[ C^{-1/2}\Sigma C^{-1/2}{\cal P}_1\right] -\tr\left[ C^{-1/2}\Sigma C^{-1/2}{\cal P}_0\right] \,, 
\end{align}
where ${\cal P}_1$ is simply the Bayesian extension of the projector ${\cal P}$ defined in eq.~\eqref{eq:proj} and ${\cal P}_0$ has a similar definition, see eqs.~\eqref{eq:P0} and~\eqref{eq:P1} and App.~\ref{sec:AppB} for the full derivation. Note that the last term appearing on the r.h.s. of eq.~\eqref{eq:RIBAIC2} encodes the residual bias induced by a misspecification of the priors. It reads
\begin{align}
\label{eq:epsilon}
    \varepsilon(\tilde\bstheta) = \left(\tilde\bstheta-\bstheta_{\rm MAP}^\text{\tiny{$C$}}\right)^T {\cal Q}^{-1} \left(\tilde\bstheta-\bstheta_{\rm MAP}^\text{\tiny{$C$}}\right) \,, 
\end{align}
where the matrix ${\cal Q}^{-1}$ is defined in eq.~\eqref{eq:Q}.
As already remarked, the underlying assumption of the entire derivation is that the model, and the priors in a Bayesian framework, are correctly specified. In this scenario, $\varepsilon(\tilde\bstheta)$ provides a small correction and the overall bias in the Bayesian framework is expected to be numerically similar to the one in eq.~\eqref{eq:bias_gAIC_2}.

As a final remark, note that in the limit of non-informative priors eq.~\eqref{eq:RIBAIC2} correctly reproduces the RIAIC in eq.~\eqref{eq:gAIC}, while it reproduces the standard AIC if the correlated limit is taken as well.
Conversely, if only the correlated limit is taken, eq.~\eqref{eq:RIBAIC2} reproduces the expression of the BAIC derived in Ref.~\cite{Neil:2022joj} up to small corrections induced by the difference between the MLE and the MAP estimators, since the expression in Ref.~\cite{Neil:2022joj}, by construction, assumes the equivalence of the two estimators.
Similarly to eq.~\eqref{eq:gAIC_full}, the RIBAIC presented in this section can be readily generalized to dataset selection problems by including the standard term for the non-fitted dataset.

\section{Numerical example}
\label{sec:example}
In this section we numerically test the Information Criteria discussed in the previous sections
on data obtained from lattice QCD simulations. In particular, we compare the RIAIC and the $\overline{\operatorname{AIC}}$
and use the value of the AIC obtained from the corresponding fully-correlated fit as a benchmark.

\subsection{Setup}
\label{sec:setup}
The Information Criteria are tested on the zero-momentum single-nucleon correlation function computed on the D251 ensemble generated by the CLS initiative and already presented in Refs.~\cite{BaSc:2026fdy,BaryonScattering:2026voj,Baeza-Ballesteros:2026qxo}.
The correlator has been computed within the distillation framework on 32 timeslices on $N_{\rm cfg} = 209$ independent gauge field configurations; further details
on the ensemble and the number of eigenvectors of the covariant three-dimensional Laplacian operator used are given in Ref.~\cite{BaryonScattering:2026voj}.
The subsequent statistical analysis employs the jackknife resampling method. In all the fits we keep $t_{\rm max}/a = 31$ fixed and vary the size of the
fitted dataset by varying $t_{\rm min}/a$, i.e. the first timeslice included in the fit.

As fit ansatz we use a multi-state model parameterized as
\begin{align}
\label{eq:model}
\eta_{\text{\tiny $n$}} (t) = A_0 e^{-E_0 t} \left( 1+ \sum_{j=1}^{n-1} |r_j|^2 e^{-\Delta E_{j} t} \right) \, .
\end{align}
In the following, a model is considered to be correctly specified if the $p$-value of the corresponding fit is larger than a threshold, which is set a priori to the value
$0.05$; if the fit is regularized, the $p$-value is computed as in Ref.~\cite{Bruno:2022mfy}. Fits which do not satisfy this requirement are therefore discarded.

For each regularized fit, the sample covariance matrix $\Sigma$ of the fitted dataset is replaced by its regularized version $C$. The regularized
covariance matrix is obtained through an SVD cut which acts on the smallest eigenvalues of $\Sigma$. In particular, eigenvalues which 
are smaller than $\lambda_{\rm max}/\bar\kappa$, where $\lambda_{\rm max}$ is the largest eigenvalue of $\Sigma$ and $\bar{\kappa}$ is
a threshold on the condition number of $C$ which is chosen a priori, are set to be equal to $\lambda_{\rm max}/\bar\kappa$, such that $\kappa(C)=\bar\kappa$.
Conversely, if $\kappa(\Sigma)\le\bar\kappa$, no SVD cut is applied to $\Sigma$. Note that we work under the same assumptions used in the derivations
of the previous sections; therefore, no regularization is applied to the covariance matrix of the non-fitted dataset $\Sigma_{\text{\tiny C}}$.
In Sec.~\ref{sec:regularization} the threshold $\bar\kappa$ is varied in order to probe the regularization dependence of the criteria,
while in the subsequent sections it is fixed a priori to the condition number of 
the covariance matrix of the dataset in $t\in[25a,31a]$, i.e.\ $\bar\kappa=\kappa\left(\Sigma_{[25a,31a]}\right)$,
which corresponds to the smallest fitted dataset considered in Sec.~\ref{sec:dataset_and_model}.

The comparison among the different Information Criteria is carried out by employing the expressions
in eq.~\eqref{eq:gAIC_full2} for the RIAIC, eq.~\eqref{eq:uAIC_full2} for the $\overline{\operatorname{AIC}}$, 
and eq.~\eqref{eq:AIC_full2} for the standard AIC, with block diagonal covariance matrices, i.e. 
by neglecting correlations between the fitted and non-fitted subsets, as described in the previous sections.

For the chosen datasets, the resulting covariance matrices, although ill-conditioned, are not singular and fully-correlated
fits are still feasible. This allows us to use the AIC obtained from the corresponding correlated fits as a benchmark
to test whether the RIAIC reproduces its outcome when regularized fits are performed on the same dataset.
In realistic scenarios, e.g. on larger datasets, where fully-correlated fits are not viable, the RIAIC then
provides a regularization-independent alternative where the standard correlated AIC cannot be computed.
Since the RIAIC is expected to be, on average, independent of the chosen regularization, small deviations 
compared to the AIC are expected if the overall quality of the fit is the same and if the fit parameters can
still be determined reliably. Conversely, larger deviations are expected to appear for the  $\overline{\operatorname{AIC}}$
due to its inherent regularization dependence. In the subsequent analysis we, therefore, compare different criteria
through the differences $\Delta_{\text{\tiny{IC}}}={\rm IC}-{\rm AIC}$ and through the model and subset they select.

The tests are organized as follows. In Sec.~\ref{sec:regularization} we study the regularization dependence of the
criteria for a fixed dataset and model. In Secs.~\ref{sec:dataset} and~\ref{sec:dataset_and_model} the criteria are applied to
dataset selection and to combined dataset and model selection, respectively. 
Finally, in Sec.~\ref{sec:bayesian} we consider
fits with Bayesian priors.

\subsection{Regularization dependence}
\label{sec:regularization}
As a first numerical test we study the regularization dependence of the RIAIC and of the
$\overline{\operatorname{AIC}}$ for a fixed dataset and model. We fit the single-nucleon
correlator with a single-exponential model, $\eta_1$. As previously mentioned,
the benchmark for the comparison is provided by the AIC obtained from correlated fits on the dataset.
To this aim, we identify the optimal subset by performing fully-correlated fits with $t_{\rm max}=31a$
and different choices of $t_{\rm min}$. Fits with $p$-value lower than $0.05$ are discarded and the smallest
AIC identifies $t\in[22a,31a]$ as the optimal window for the chosen single-exponential model.

On the same data subset we apply several SVD cuts to its covariance matrix $\Sigma$,
such that $\kappa(\Sigma)/\kappa(C)$ ranges from $1$ to $\sim 4\times10^3$,
and perform the corresponding regularized fits. The quality of the fit in this case
is assessed as in Ref.~\cite{Bruno:2022mfy} and we employ the same threshold used for the 
correlated case.

\begin{figure}
    \centering
    \includegraphics[width=\linewidth]{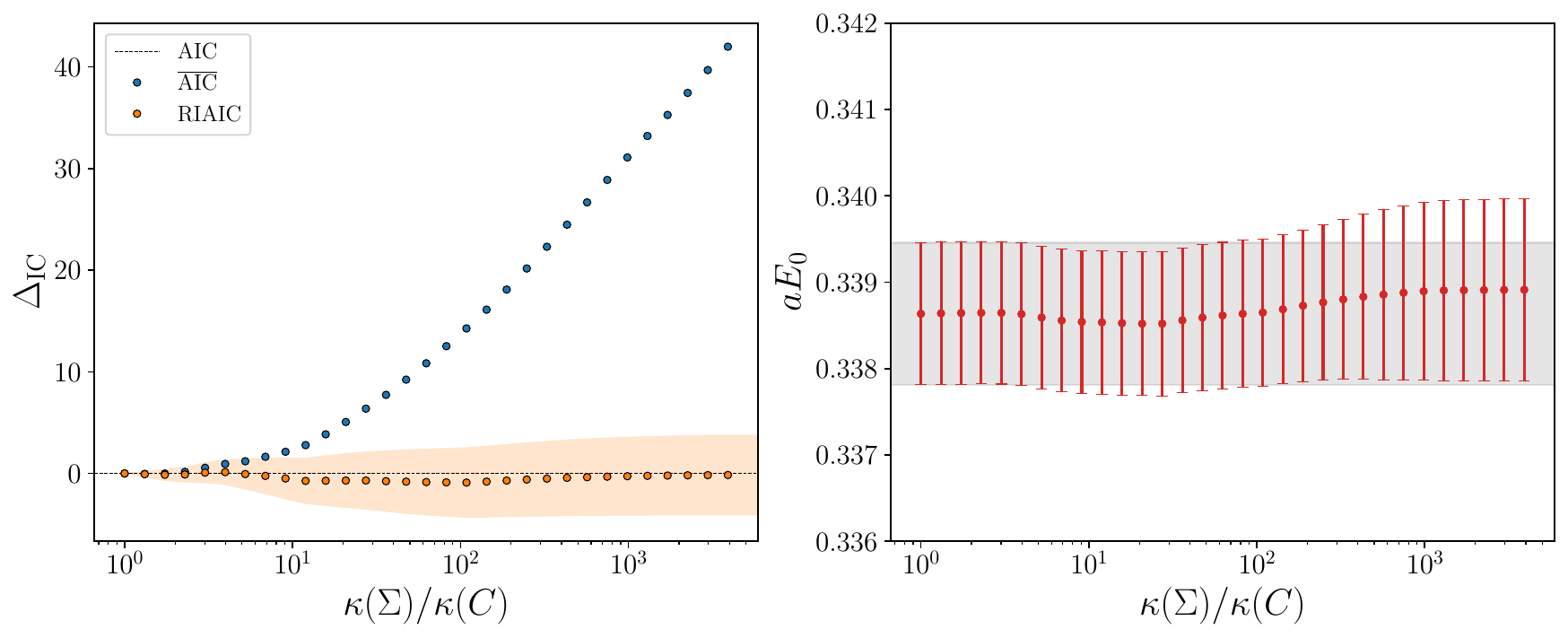}
    \caption{Left panel: deviations of the ICs under study compared to the standard AIC as a function of the regularization expressed in terms of $\kappa(\Sigma)/\kappa(C)$. Orange points correspond to the RIAIC, while blue points correspond to the $\overline{\operatorname{AIC}}$. The orange band is the error computed as in App.~\ref{sec:AppC}. Right panel: ground-state determinations for each fit as a function of $\kappa(\Sigma)/\kappa(C)$. The horizontal gray band corresponds to the result obtained from a correlated fit. The fitted dataset has been kept fixed to the window $t\in[22a,31a]$ and the model is the single-exponential one.}
    \label{fig:ICs}
\end{figure}

In the left panel of Figure~\ref{fig:ICs} we show $\Delta_{\text{\tiny{IC}}}={\rm IC}-{\rm AIC}$
as a function of $\kappa(\Sigma)/\kappa(C)$. For $\kappa(\Sigma)/\kappa(C)=1$ both
the $\overline{\operatorname{AIC}}$ and the RIAIC reproduce the standard AIC, as expected.
When stronger regularizations are applied, the $\overline{\operatorname{AIC}}$ increasingly
deviates from the AIC, up to $\Delta_{\text{\tiny{IC}}}\simeq 42$ for the strongest regularization considered,
and the deviation is expected to grow further for even stronger regularizations.
This behavior is not surprising; indeed, the $\overline{\operatorname{AIC}}$ in eq.~\eqref{eq:UAIC1}
is an unbiased estimator of the regularization-dependent $\Delta(C \mid\bstheta,\eta)$,
which measures the distance between the data-generating probability distribution, containing $\Sigma$,
and the fitted regularized distribution, containing $C$. The SVD cut
acts as a regulator by raising the smallest eigenvalues of $\Sigma$; this 
implies that $\log\det C > \log\det\Sigma$, and, therefore,
the $\overline{\operatorname{AIC}}$ is dominated by this term as the regularization becomes stronger, 
which is the behavior observed in Figure~\ref{fig:ICs}.

On the other hand, the RIAIC is an estimator of the correlated $\Delta(\Sigma \mid\bstheta,\eta)$,
which is, by definition, regularization-independent, and only contains the determinant of the covariance matrix $\Sigma$.
As a consequence, its value remains stable across the entire range of regularizations explored and exhibits a deviation
$\Delta_{\text{\tiny{IC}}}$ which is always below one unit and compatible with zero within the error
computed as in eq.~\eqref{eq:variance}, suggesting that such deviation is mainly due to statistical fluctuations
of $\chi^2_C$ around its expectation value.

In the right panel of Figure~\ref{fig:ICs} we show the obtained
ground-state energy estimates. Over the whole range of $\kappa(\Sigma)/\kappa(C)$,
the correlated difference between the ground-state energy from a regularized fit
and the one obtained from the corresponding correlated fit
is compatible with zero within $\sim 1$ standard deviation, and its statistical error increases
by at most $\sim 30\%$. In particular, for $\kappa(\Sigma)/\kappa(C)\lesssim 10^2$, the standard
deviation of the estimated ground-state energy obtained from the regularized fit differs by less
than $5\%$ from the correlated result.
The latter result, together with the fact that deviations between the RIAIC and the AIC differ from zero by less than
one standard deviation, confirms that, for moderate regularizations, the information removed by the SVD cut does not affect the overall 
final estimate of the ground-state energy.
Conversely, at the same values of $\kappa(\Sigma)/\kappa(C)$, 
$\Delta_{\text{\tiny{IC}}}$ is $O(10)$ for the $\overline{\operatorname{AIC}}$ and may lead
to the wrong identification of the optimal subset or model, or, in general, affects the
final estimate in model averaging procedures.

\subsection{Dataset selection}
\label{sec:dataset}
We fit the single-nucleon correlator used in the previous section with a two-exponential
model $\eta_2$, keeping $t_{\rm max}=31a$ fixed and varying $t_{\rm min}$.
As before, fits with a $p$-value lower than $0.05$ are discarded from the analysis.
In addition, we exclude from the analysis fits where excited-state contamination
cannot be resolved, i.e. when the associated parameters are compatible with zero within one
standard deviation. For both correlated and regularized fits, this leads to restricting 
the analysis to fit windows with $8a\le t_{\rm min}\le 15a$. The benchmark optimal
fit range is identified through the AIC of fully-correlated fits and corresponds
to the window starting at $t_{\rm min}=10a$.

The $\overline{\operatorname{AIC}}$ and the RIAIC are tested in the context
of data subset selection, by repeating the same fits with the regularization described
in Sec.~\ref{sec:setup}, i.e. with threshold on the condition number of $C$
given by $\bar\kappa=\kappa\left(\Sigma_{[25a,31a]}\right)$.
The strength of the regularization, $\kappa(\Sigma)/\kappa(C)$,
increases by lowering $t_{\rm min}$ from $\sim 10^2$ at $t_{\rm min} = 15a$
to $\sim 4\times10^3$ at $t_{\rm min}=8a$,
i.e. within the range explored in Sec.~\ref{sec:regularization}.

\begin{figure}[ht]
    \centering
    \includegraphics[width=\linewidth]{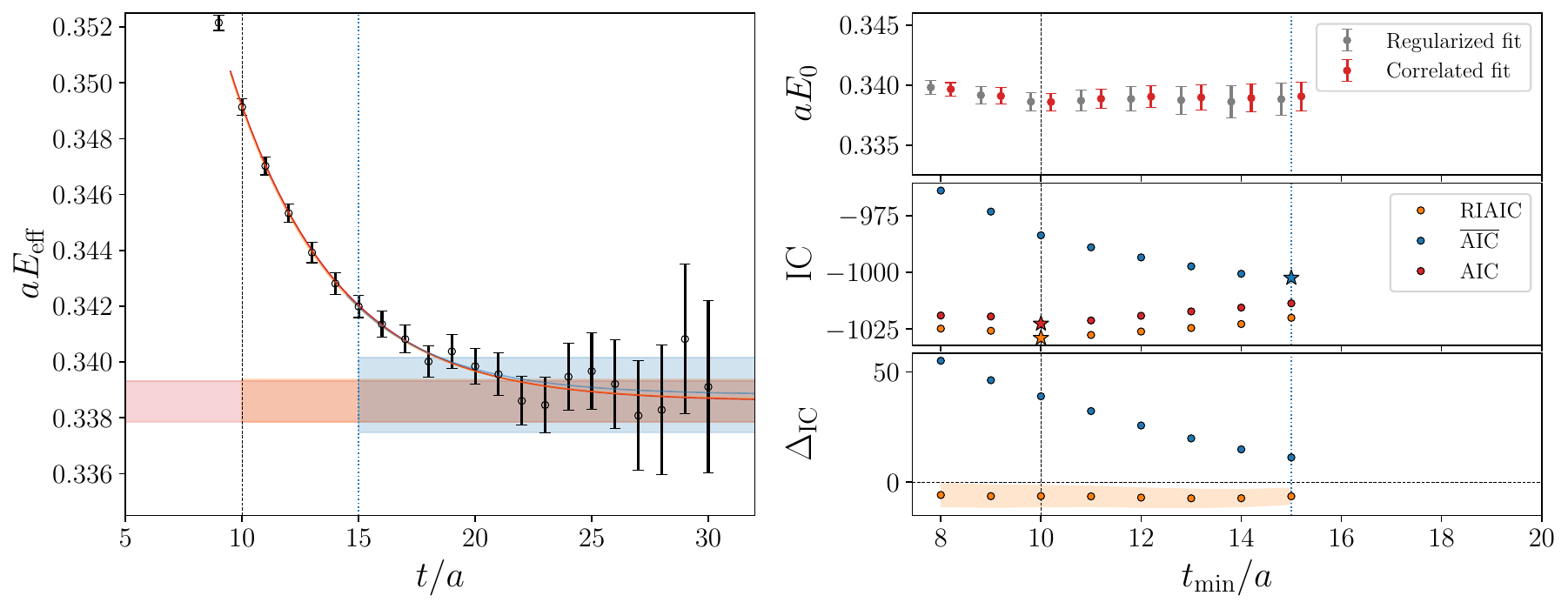}
    \caption{Left panel: the effective energy of the single-nucleon correlator. The lines and bands represent the optimal double-exponential model and the estimated ground-state energy selected by the AIC in a correlated fit (red) and by the RIAIC (orange) and the $\overline{\operatorname{AIC}}$ (blue) in a regularized fit. Upper right panel: ground-state energies obtained for several choices of $t_{\rm min}$ with correlated (red) and regularized (gray) fits. Middle right panel: $t_{\rm min}$ plot with the values of the AIC (red), RIAIC (orange) and $\overline{\operatorname{AIC}}$ (blue). Stars mark the optimal $t_{\rm min}$ according to each criterion. The vertical lines represent the optimal $t_{\rm min}$ identified using the AIC and the RIAIC (dashed line) and the $\overline{\operatorname{AIC}}$ (dotted line). Lower right panel: the deviation $\Delta_{\text{\tiny{IC}}}$ of each IC from the AIC; the orange band is the error associated with the difference between the RIAIC and the AIC; see App.~\ref{sec:AppC}.}
    \label{fig:ICs-dataset}
\end{figure}

The effective energy of the single-nucleon correlator is shown in the left panel of
Figure~\ref{fig:ICs-dataset}, together with the optimal correlated double-exponential
fit starting at $t_{\rm min}=10a$ and the estimates of the ground-state energy selected by the ICs under study.

In the middle right panel we show the values of the three Information Criteria under study
for each $t_{\rm min}$. Red points correspond to the AIC obtained from correlated fits,
used as a benchmark, while orange and blue points represent the RIAIC and the
$\overline{\operatorname{AIC}}$, respectively. 
The RIAIC exhibits deviations of a few
units from the AIC over the whole range of $t_{\rm min}$.
An interpretation of such deviations is given by the lower panel on the right,
where we explicitly show the difference $\Delta_{\text{\tiny{IC}}}$ for both the RIAIC and the $\overline{\operatorname{AIC}}$.
The orange band is the error associated with the deviations of the RIAIC relative to the standard AIC, computed from eq.~\eqref{eq:variance}.
In general, the difference between the RIAIC and the AIC deviates from zero by less than $1.8$ standard deviations, and
by $1.2$ standard deviations for $t_{\rm min} = 10a$, which is identified as the optimal one by the RIAIC.
The RIAIC exhibits a systematic shift to smaller values compared to the AIC. This
behavior is understood from the fact that the values of $\Delta_{\text{\tiny{RIAIC}}}$ obtained on
different fit windows are strongly correlated, with correlation coefficients $\rho\in[0.79,0.99]$
computed from eq.~\eqref{eq:cov_windows}.

Ground-state energies obtained from regularized and correlated fits are
shown in the upper right panel of Figure~\ref{fig:ICs-dataset}. These are compatible
for all $t_{\rm min}$ and, therefore, deviations between the AIC and the RIAIC can be traced back
to fluctuations of $\chi^2_C$ around its expectation value, as discussed in Sec.~\ref{sec:regularization}.
Remarkably, since such deviations are approximately constant in $t_{\rm min}$, they do not affect the selection and
the RIAIC identifies the same optimal window, starting
at $t_{\rm min}=10a$, obtained by applying the AIC to fully-correlated fits.

Conversely, $\Delta_{\text{\tiny{IC}}}$ for the $\overline{\operatorname{AIC}}$ increases monotonically from $t_{\rm min}=15a$ to $t_{\rm min}=8a$. As discussed in Sec.~\ref{sec:regularization}, this behavior is due to the fact that the $\overline{\operatorname{AIC}}$ is an estimator of the regularization-dependent $\Delta(C \mid\bstheta,\eta)$.
As a consequence, the $\overline{\operatorname{AIC}}$ selects the largest $t_{\rm min}$ among those 
which yield a valid fit, i.e. it promotes subsets which are less regularized and, therefore, loses 
the usual trade-off between quality of the fit and complexity of the description.

As a final remark, the identification of different optimal data subsets affects the estimate
of the ground-state energy as shown in the upper right panel of Figure~\ref{fig:ICs-dataset}.
The correlated difference between the determinations based on the RIAIC and the AIC is compatible with
zero within less than 0.2 standard deviations and the two ground-state energy estimates have similar
statistical errors. In contrast, the determination obtained with the $\overline{\operatorname{AIC}}$,
despite being compatible with the estimate based on the AIC, has a statistical uncertainty which
is almost twice as large.

\subsection{Dataset and model selection}
\label{sec:dataset_and_model}
In this section we test one- and two-exponential models $\eta_1$ and $\eta_2$ in eq.~\eqref{eq:model} 
for different choices of $t_{\rm min}$
to identify the optimal subset-model description of the data, based
on the Information Criteria under study.
As before, we use as benchmark the results based on the AIC obtained from correlated fits 
and we discard or accept models based on the same criteria discussed in Sec.~\ref{sec:dataset}.
In the correlated analysis, the double-exponential model is preferred for $8a\le t_{\rm min}\le 15a$
and the single-exponential one for $18a\le t_{\rm min}\le 25a$.
For $t_{\rm min}=16a,17a$ no model is accepted based on our criteria. The reason
is that, for such windows, single-exponential models are discarded based on their poor quality
due to the residual excited-state contamination at low $t_{\rm min}$,
while double-exponential models cannot clearly resolve
the contribution from the first excited state.
As in the previous section, the correlated analysis based on the AIC
identifies the double-exponential model starting at
$t_{\rm min} = 10a$ as the optimal subset-model description.
We then repeat the same analysis by performing regularized fits on each dataset in order
to compare the outcome of the $\overline{\operatorname{AIC}}$ and RIAIC. As before, we
regularize the covariance matrix of a given dataset if its condition number is larger than
a given threshold set to $\bar\kappa = \kappa\left(\Sigma_{[25a,31a]}\right)$. For
each $t_{\rm min}$ we only select the model with the lowest IC.

\begin{figure}
    \centering
    \includegraphics[width=\linewidth]{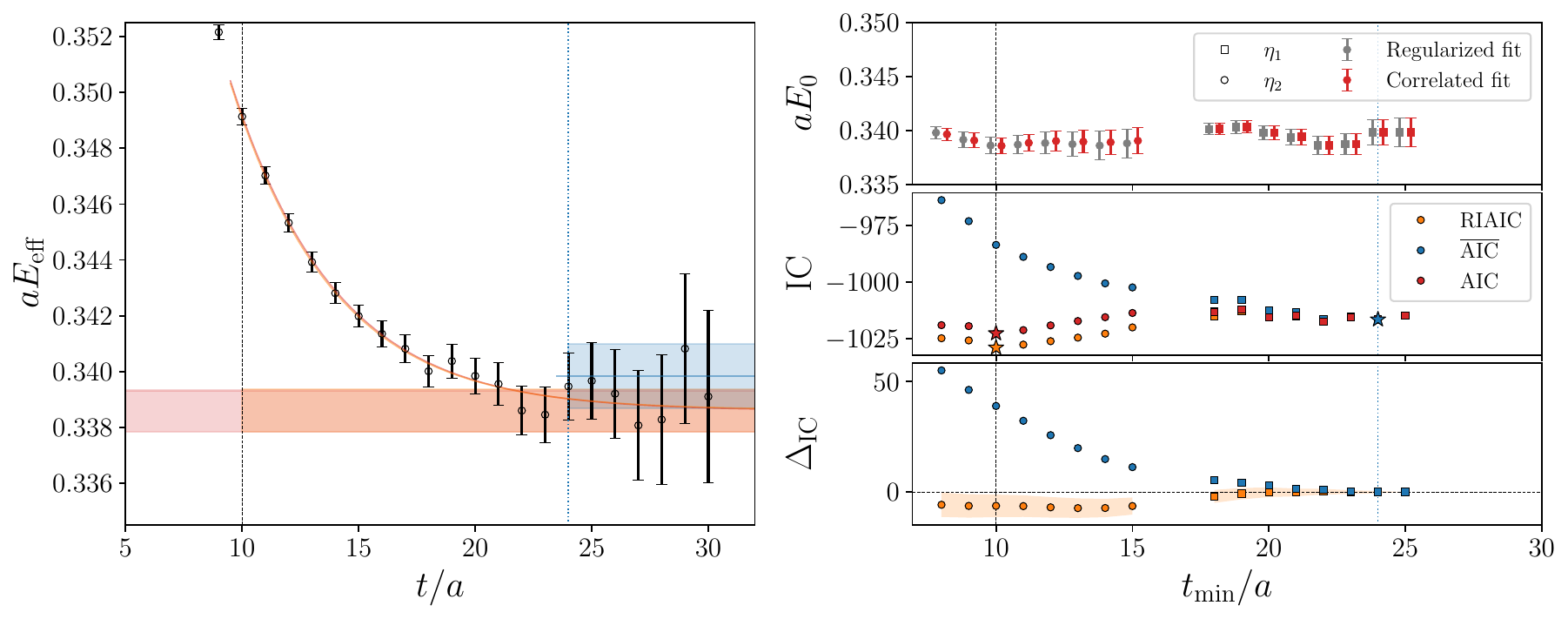}
    \caption{As in Figure~\ref{fig:ICs-dataset}, but for simultaneous dataset and model selection. Squares and circles
denote single- ($\eta_1$) and double-exponential ($\eta_2$) fits, respectively.
For each $t_{\rm min}$, only the model with the lowest value of the corresponding
Information Criterion is shown; the vertical lines mark the optimal
subset-model description identified by the AIC and the RIAIC (dashed line) and by the $\overline{\operatorname{AIC}}$
(dotted line).}
    \label{fig:ICs-dataset-model}
\end{figure}

In the middle right panel of Figure~\ref{fig:ICs-dataset-model} we show the values of the three
Information Criteria for each $t_{\rm min}$. 
Let us stress that, for each $t_{\rm min}$, both the RIAIC and the $\overline{\operatorname{AIC}}$
identify the same optimal model obtained from the benchmark analysis based on the standard AIC.
For $8a\le t_{\rm min}\le 15a$, the double-exponential model is preferred,
and the new analysis reproduces the results seen in Sec.~\ref{sec:dataset}, with the 
RIAIC deviating from the AIC by a few units and the $\overline{\operatorname{AIC}}$
deviating by up to $\sim 55$ units from the AIC. For $18a\le t_{\rm min}\le 25a$ 
all the ICs identify the single-exponential model as the optimal one and deviations 
from the AIC are smaller both for the RIAIC and the $\overline{\operatorname{AIC}}$ due 
to the much milder regularization applied to the covariance matrix of the fitted dataset.
The limiting case is achieved for $t_{\rm min}=25a$, where no regularization is applied
and the three ICs coincide.
As a consequence, the RIAIC identifies the same subset-model description obtained by applying the AIC
to correlated fits, i.e. the double-exponential model starting at $t_{\rm min}=10a$, whereas the
$\overline{\operatorname{AIC}}$ selects the single-exponential model starting at $t_{\rm min}=24a$.
As discussed in Sec.~\ref{sec:regularization}, this result is mainly due to the regularization dependence
of the $\overline{\operatorname{AIC}}$ which penalizes regularized subsets with low $t_{\rm min}$ and
promotes fit windows with larger values of $t_{\rm min}$ where the regularization has smaller effects, 
i.e. the regularized covariance matrix $C$ is closer to the one with intact correlations.

The lower right panel shows that, for $8a\le t_{\rm min}\le 15a$, the deviation $\Delta_{\text{\tiny{IC}}}$ of the RIAIC
coincides with the one discussed in Sec.~\ref{sec:dataset}, while for $t_{\rm min} \ge 18a$ it is compatible
with zero within less than 0.7 standard deviations. As expected, the statistical error shrinks as $t_{\rm min}$ increases and vanishes at
$t_{\rm min}=25a$, where $C=\Sigma$; see eq.~\eqref{eq:variance}.

The impact of the different selections on the ground-state energy is shown in the upper right panel of
Figure~\ref{fig:ICs-dataset-model}. 
Regularized and correlated fits yield compatible results for all $t_{\rm min}$ both for single-
and double-exponential models.
The determination based on the RIAIC agrees with the one based on the AIC: their correlated difference is
compatible with zero within less than 0.2 standard deviations and the two estimates have similar statistical errors.
In contrast, the determination based on the $\overline{\operatorname{AIC}}$, obtained from the single-exponential fit
starting at $t_{\rm min}=24a$, results in a correlated difference which is $\sim 1.3$ standard deviations
away from zero and a statistical uncertainty which is more than $1.5$ times larger.

These results show that, when regularizations are applied to the covariance matrix,
the $\overline{\operatorname{AIC}}$ may lead to a different subset-model selection.
On the other hand, the RIAIC, whose bias accounts for the regularization of the covariance matrix,
is expected to reproduce, on average, the
selection obtained by applying the AIC to correlated fits.

As a final remark, note that the small deviations between the RIAIC and the 
AIC may, in principle, lead to different results if the ICs are used as weights
in model-averaging procedures. However, these differences are expected to be
even larger if the $\overline{\operatorname{AIC}}$ is used instead.

\subsection{Bayesian fits and prior dependence}
\label{sec:bayesian}
In this section we test the RIBAIC defined in eq.~\eqref{eq:RIBAIC2} on Bayesian fits to
the same single-nucleon correlator analyzed in the previous numerical tests.
As discussed in Sec.~\ref{sec:bayes_extension}, in the limit of non-informative priors,
the RIBAIC is expected to reproduce the RIAIC in eq.~\eqref{eq:gAIC} and, if the correlated
limit is taken as well, i.e. $C=\Sigma$, it is expected to reproduce the standard AIC.
To this aim, we keep fixed both the model and the fitted dataset based on the results
of Secs.~\ref{sec:dataset} and~\ref{sec:dataset_and_model}, i.e. we fit
the dataset starting at $t_{\rm min} = 10a$ with the double-exponential model, and we vary
the width of the priors.

We use Gaussian priors for all the model parameters, i.e. $n_\pi = n_\eta = 4$ with diagonal
prior covariance matrix $C_\theta$ and priors given by
\begin{align}
E_0 \sim {\cal N} \left( \hat{E}, \lambda \hat{E}  \right) ,\quad A_0 \sim {\cal N}\left( \hat{A}, \lambda \hat{A}  \right) ,\quad \Delta E_1 \sim {\cal N}\left(2m_\pi, 2\lambda m_\pi\right) , \quad r_1 \sim {\cal N}\left(1, \lambda \right) , 
\end{align}
where the second argument in ${\cal N}(\cdot,\cdot)$ here denotes the standard deviation. 
In the priors above, $\hat{E}$ and $\hat{A}$ are estimated from the effective energy at $t=23a$ by assuming
a single-state saturation at the selected timeslice and $am_\pi = 0.09$; for a more precise
estimate of the pion mass on this ensemble see Table II of Ref.~\cite{BaryonScattering:2026voj} and Table 4 of Ref.~\cite{Kuberski:2024pms}.
The parameter $\lambda$ is varied between $10^{-3}$ and $10^{2}$ to probe the 
effects of constraining and non-informative priors.
For each selection of priors, we perform both fully-correlated and regularized fits with 
covariance matrix $\Sigma$ and $C$ respectively, where $C$ is obtained by applying an SVD
cut such that its condition number is equal to $\bar\kappa=\kappa\left(\Sigma_{[25a,31a]}\right)$.
The quality of the Bayesian fit is assessed through the $p$-value associated with the augmented chi-square
which includes the contribution of the priors~\cite{Lepage:2001ym}.

\begin{figure}[ht]
    \centering
    \includegraphics[width=\linewidth]{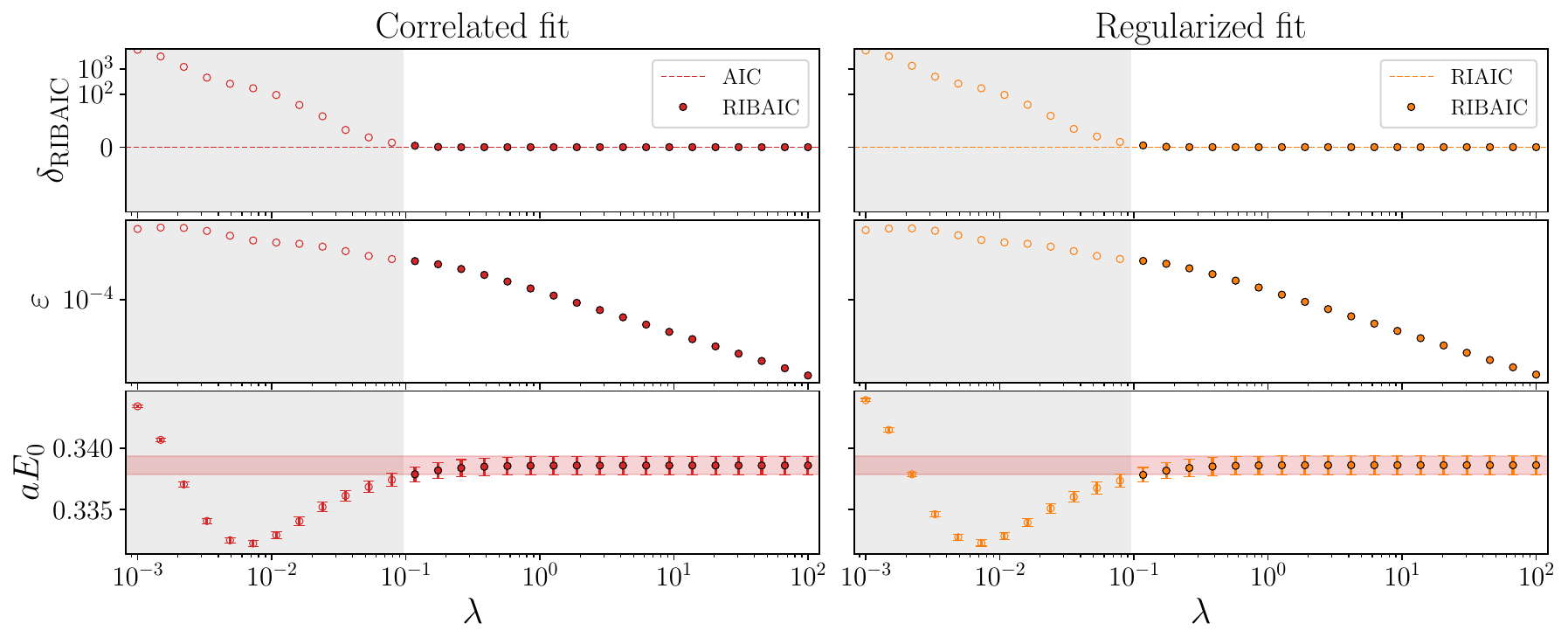}
    \caption{Results from Bayesian fits to the single-nucleon correlator with the double-exponential model $\eta_2$ on the window $t\in[10a,31a]$,
as a function of the relative width $\lambda$ of the priors. Left (right) column: correlated (regularized) fits. Upper panels: deviation
of the RIBAIC from the AIC (RIAIC) of correlated (regularized) frequentist fits, $\delta_{\text{\tiny{RIBAIC}}}$; the dashed lines mark the AIC (left) and the RIAIC (right),
i.e. the expected limits for non-informative priors. Middle panels: the term $\varepsilon(\tilde\bstheta)$ in eq.~\eqref{eq:epsilon}.
Lower panels: ground-state energy estimates. The red band corresponds to the result of the correlated frequentist fit.
Open symbols in the gray regions denote fits with Bayesian $p$-value lower than $0.05$.}
\label{fig:ICs-Bayesian}
\end{figure}

In Figure~\ref{fig:ICs-Bayesian} we show the results of the analysis for both the correlated and 
the regularized fits. 
For $\lambda\gtrsim 10^{-1}$ all the fits have $p$-value larger than the selected threshold
and the RIBAIC approaches the expected non-informative limit, i.e. the standard AIC in the
case of fully-correlated fits and the RIAIC in the regularized case.
Note that, in contrast to $\Delta_{\text{\tiny{IC}}}$ in the previous numerical tests,
here $\delta_{\text{\tiny{RIBAIC}}}$ denotes the difference between the RIBAIC and the corresponding
expected limit for $\lambda\to\infty$.
In particular, for $\lambda \gtrsim 0.8$ the deviations of the RIBAIC from the expected non-informative limit
are already below $10^{-3}$ and the ground-state energy estimates reported in the lower panels are stable 
and with a statistical error similar to the one extracted from the correlated fit.
In this regime, since the MAP and the MLE estimators are numerically close, 
the correlated RIBAIC (left column) also reproduces the BAIC of Ref.~\cite{Neil:2022joj} up to small corrections.
In the middle panels we show the term $\varepsilon(\tilde\bstheta)$ as a function of $\lambda$ for both 
the correlated (left) and the regularized (right) fits. As expected, when the width of the priors
increases, $\varepsilon(\tilde\bstheta)$ monotonically decreases and, for priors with $\lambda\gtrsim 1$,
it is already of the order of $10^{-4}$.

For $\lambda\lesssim 10^{-1}$, fits are prior dominated and do not provide stable results for the fit parameters.
The ground-state energy estimate approaches the prior means as $\lambda\to 0$.
Similarly, the RIBAIC increases and, in the same limit, is expected to saturate to the value of $\chi^2_C$
evaluated at the prior means. Note that, even though fits in this region are discarded based on their $p$-value,
the value of $\varepsilon(\tilde\bstheta)$ remains relatively small and does not signal the misspecification
of the priors. The main reason, as discussed in App.~\ref{sec:AppB}, is that the MAP estimator is biased if
priors are misspecified and therefore the substitution $\bstheta_0 \approx \bstheta_{\rm MAP}^\text{\tiny{$C$}}$ is no longer
justified and may lead to an unreliable estimate of the value of $\varepsilon(\tilde\bstheta)$.

These results show that, for moderately wide priors, the RIBAIC of a Bayesian fit and the RIAIC of the
corresponding frequentist fit on the same dataset are numerically equivalent. 
This is due to the fact that, under the assumption that both model and priors are correctly
specified, these criteria are estimators of the same correlated
quantity, independently of any type of regularization. As a consequence, they can be used
in a unified framework, in which frequentist and Bayesian fits, possibly with different regularizations,
are compared in model and dataset selection or combined in model-averaging procedures.

\section{Summary}
In this work, we have derived an extension of the Akaike Information Criterion which
allows one to tackle model and data subset selection problems when regularized fits are performed on correlated data.
The main result is represented by the RIAIC and is reported in eq.~\eqref{eq:gAIC} for standard frequentist fits. The RIAIC
allows the comparison between models and subsets when regularizations are applied to the covariance matrix
of the fitted dataset. The generalization to maximum a posteriori fits, i.e. the RIBAIC, has been derived
as well and is reported in eq.~\eqref{eq:RIBAIC2}. In Sec.~\ref{sec:example}, we tested
the numerical and statistical properties of the RIAIC and of the RIBAIC. We found that, while
the naive generalization of the AIC to the case of regularized fits, which throughout this work we called the $\overline{\operatorname{AIC}}$,
systematically favors less regularized datasets, the RIAIC deviates from the AIC of the corresponding correlated fit only by a few units, consistently
with the statistical fluctuations in $\chi^2_C$ around its expectation value; see App.~\ref{sec:AppC}. 
In the numerical tests that we carried out, this led to the identification of the same optimal model and data subset obtained with the AIC, as expected
since the two criteria have the same expectation value.

The RIAIC and the RIBAIC were obtained by defining the Information Criterion as an unbiased estimator of
the correlated model-dependent contribution to the KL divergence. Equivalently, the expectation values of
the RIAIC and of the RIBAIC are equal to the expectation value of the standard, correlated Akaike
Information Criterion. By taking the correlated limit, both the RIAIC and the RIBAIC reproduce 
standard results in the literature, namely the AIC~\cite{Akaike1,Akaike2} and, if the MAP and the MLE estimators are numerically close,
the BAIC~\cite{Neil:2022joj}, respectively.

These criteria apply in the most general setting, where both regularizations of the covariance
matrix and priors are included in the fit. We interpret the use of regularizations at the level of
the covariance matrix and the inclusion of priors as a modification of the total amount of information encoded
in the data we aim to model. In the first case, regularizing the covariance matrix has the role of removing
poorly-estimated information. Conversely, the inclusion of priors provides additional information
which effectively acts as a regulator of the fit. In both cases the regularization is properly absorbed
into the bias of the Information Criteria in eqs.~\eqref{eq:gAIC} and~\eqref{eq:RIBAIC2}, leading
to an overall result which is, on average, independent of the chosen regularization.

The estimators that have been derived are obtained starting from the Akaike Information Criterion.
Nonetheless, the generalization to other Information-Theoretic Criteria is possible.

The application of these results encompasses not only data subset and model selection, but also 
model averaging procedures as detailed in Ref.~\cite{Jay:2020jkz}.
In particular, as shown in Sec.~\ref{sec:bayesian}, for moderately wide priors,
the RIBAIC reproduces the numerical value of the RIAIC. 
As a consequence, frequentist and Bayesian fits performed
on the same dataset, with or without regularization, can be compared
and combined within a unified framework, either in model and dataset selection or in model averaging procedures.

\section*{Acknowledgements}
I thank my co-authors of Refs.~\cite{BaryonScattering:2026voj,Baeza-Ballesteros:2026qxo} and colleagues of the Baryon Scattering (BaSc) Collaboration for sharing the single-nucleon
correlator used in this work, in particular F.~Romero-L\'opez for computing perambulators on the D251 ensemble and J.R.~Green for the Wick contractions.
I would like to thank J.~Bulava for comments on an early version of the manuscript and A.~Bai\~ao Raposo, M.~Bruno and M.~Ginesi for useful discussions on 
the topic of this work. My work is supported by the European Research Council (ERC) consolidator grant StrangeScatt-101088506. Part of the analysis in Sec.~\ref{sec:example}
was performed on Elysium, the HPC cluster of the Ruhr-Universit\"at Bochum, subsidized by the DFG(INST213/1055-1). The analysis of Sec.~\ref{sec:bayesian}
was performed using the \texttt{gvar}~\cite{gvar} and \texttt{lsqfit}~\cite{lsqfit} libraries.

\appendix
\section{RIAIC bias correction}
\label{sec:AppA}
Recall the expression of the bias appearing in the regularization-independent definition of the AIC in eq.~\eqref{eq:bias_gAIC}
\begin{align}
{\cal B}(\bsy, C,\eta) \equiv \mathbb{E} \Big[ 2\log {\cal L}(\bsy,C \mid \bstheta_{\rm MLE}^\text{\tiny{$C$}},\eta) + \Delta(\Sigma \mid\bstheta_{\rm MLE}^\text{\tiny{$\Sigma$}},\eta) \Big]  \, .
\end{align}
The derivation in this section follows the presentation in Ref.~\cite{https://doi.org/10.1002/wics.1460} and generalizes it to the case of regularized fits for the first term appearing in the bias. The same steps, applied to the definition in Sec.~\ref{sec:UAIC}, provide the expression of the bias of the $\overline{\operatorname{AIC}}$ in eq.~\eqref{eq:UAIC1}.
The first term of the equation above is readily computed by inserting the expression of the log-likelihood, which leads to
\begin{align}
\label{eq:bias1}
\mathbb{E} &\left[2\log{\cal L}(\bsy,C\mid \bstheta_{\rm MLE}^\text{\tiny{$C$}},\eta) \right] = - \mathbb{E} \left[ \chi^2_C(\bstheta_{\rm MLE}^\text{\tiny{$C$}}) + \log\det C + n_d \log(2\pi) \right] \\[.2cm]
& = -\log\det C - n_d\log(2\pi) - \tr\left[ C^{-1} \Sigma  \right] + \tr\left[ C^{-1/2} \Sigma C^{-1/2} {\cal P} \right]  ,\nonumber
\end{align}
where $\chi^2_C(\bstheta_{\rm MLE}^\text{\tiny{$C$}})$ is the regularized chi-square variable evaluated at $\bstheta_{\rm MLE}^\text{\tiny{$C$}}$. In the second line we have used the result from Ref.~\cite{Bruno:2022mfy} for the expected value of the chi-square. The projector ${\cal P}$ is defined as 
\begin{align}
\label{eq:proj}
{\cal P} = C^{-1/2} J (\bstheta_{\rm MLE}^\text{\tiny{$C$}}) H_C^{-1}(\bstheta_{\rm MLE}^\text{\tiny{$C$}}) J^T (\bstheta_{\rm MLE}^\text{\tiny{$C$}}) C^{-1/2} \, ,
\end{align}
where we have introduced the estimated $n_\eta \times n_\eta$ inverse covariance matrix of the MLE parameters, associated with the regularized matrix $C$, as
\begin{align}
\label{eq:H}
H_C (\bstheta_{\rm MLE}^\text{\tiny{$C$}}) = J^T C^{-1} J \mid_{\bstheta_{\rm MLE}^\text{\tiny{$C$}}}\, ,
\end{align}
and $J$ is the Jacobian of the model function with respect to the model parameters $\bstheta$.
As in Ref.~\cite{Bruno:2022mfy}, the unknown $\bstheta_0$ is replaced in practice by the
estimator $\bstheta_{\rm MLE}^\text{\tiny{$C$}}$.
Therefore, by introducing the effective number of parameters $n_\eta^{\rm eff}$ as in eq.~\eqref{eq:neff} and $n_d^{\rm eff} = \tr\left[ C^{-1} \Sigma  \right]$, the first term of the bias reads
\begin{align}
\mathbb{E} \left[2\log{\cal L}(\bsy,C\mid \bstheta_{\rm MLE}^\text{\tiny{$C$}},\eta) \right] =  -\log\det C - n_d\log(2\pi) - n_d^{\rm eff} + n_\eta^{\rm eff} \, .
\end{align}
The second term of the bias is evaluated at $\bstheta_{\rm MLE}^\text{\tiny{$\Sigma$}}$, i.e. the MLE parameter obtained from a \emph{correlated} fit, which is in principle not accessible if $\Sigma$ is close-to-singular. By inserting the expression in eq.~\eqref{eq:Delta}, it can be written as
\begin{align}
\label{eq:E_Delta}
\mathbb{E} \left[ \Delta(\Sigma \mid \bstheta_{\rm MLE}^\text{\tiny{$\Sigma$}},\eta) \right] = \mathbb{E} \bigg[ \left.\mathbb{E}\big[ -2\log{\cal L}(\bsy,\Sigma\mid \bstheta,\eta) \big]\right|_{\bstheta_{\rm MLE}^\text{\tiny{$\Sigma$}}} \bigg] \, .
\end{align}
By expanding the likelihood, the innermost expectation value reads
\begin{align}
\label{eq:inner_exp}
\mathbb{E} \left.\left[ -2\log{\cal L}(\bsy,\Sigma\mid \bstheta,\eta) \right]\right|_{\bstheta_{\rm MLE}^\text{\tiny{$\Sigma$}}} = \mathbb{E} \left.\left[ \chi^2_\Sigma(\bstheta) \right]\right|_{\bstheta_{\rm MLE}^\text{\tiny{$\Sigma$}}} + \log\det \Sigma + n_d \log(2\pi) \,,
\end{align}
where the expectation value of the chi-square is taken before evaluating it at the MLE parameter $\bstheta_{\rm MLE}^\text{\tiny{$\Sigma$}}$. In order to compute the expectation value appearing on the r.h.s. of eq.~\eqref{eq:inner_exp} we Taylor-expand the model function $\eta$ around the true value $\bstheta_0$. By using the standard result for the expectation value of quadratic forms and recalling that $\mathbb{E}\left[ \bsy \right] = \eta(\bstheta_0)$, the expected value on the r.h.s. reads
\begin{align}
\left. \mathbb{E}\left[ \chi^2_\Sigma(\bstheta) \right] \right|_{\bstheta_{\rm MLE}^\text{\tiny{$\Sigma$}}} = n_d + \delta\bstheta^T J^T(\bstheta_0)  \Sigma^{-1} J(\bstheta_0) \delta\bstheta \, ,
\end{align}
where we introduced $\delta \bstheta \equiv \bstheta_{\rm MLE}^\text{\tiny{$\Sigma$}} - \bstheta_0$. By combining the expression above with eqs.~\eqref{eq:E_Delta} and~\eqref{eq:inner_exp}, the second term of the bias reads
\begin{align}
\mathbb{E} \left[ \Delta(\Sigma \mid \bstheta_{\rm MLE}^\text{\tiny{$\Sigma$}},\eta) \right] = \log\det \Sigma + n_d \log(2\pi) + n_d + \mathbb{E} \left[ \delta\bstheta^T H_\Sigma(\bstheta_0) \delta\bstheta  \right] \, ,
\end{align}
where we defined $H_\Sigma$ according to eq.~\eqref{eq:H}. The remaining expectation value can be readily computed by using the condition at the minimum for $\chi^2_\Sigma(\bstheta)$, which leads to the relation
\begin{align}
\label{eq:MLE_min}
    \bstheta_{\rm MLE}^\text{\tiny{$\Sigma$}} - \bstheta_0 = H_{\Sigma}^{-1} (\bstheta_0) J^T(\bstheta_0) \Sigma^{-1} \left(\bsy-\eta(\bstheta_0)\right) \, .
\end{align}
By using this relation and the standard results for quadratic forms, the expectation value on the r.h.s. reduces to the total number of parameters of the model function $\eta$~\cite{https://doi.org/10.1002/wics.1460,Frison:2023lwb}.
Therefore the second term of the bias in eq.~\eqref{eq:bias_gAIC} can be written in the compact form
\begin{align}
\label{eq:bias2}
    \mathbb{E} \left[ \Delta(\Sigma \mid \bstheta_{\rm MLE}^\text{\tiny{$\Sigma$}},\eta) \right] = \log\det \Sigma + n_d \log(2\pi) + n_d + n_\eta \, .
\end{align}
Finally, by combining eqs.~\eqref{eq:bias1} and~\eqref{eq:bias2}, the bias takes the form in eq.~\eqref{eq:bias_gAIC_2} and the final expression for the regularization-independent criterion is the one reported in eq.~\eqref{eq:gAIC}.
To conclude this section, notice that in the correlated limit the trace involving the projector ${\cal P}$ reduces to the total number of parameters and the expression of the bias reproduces the bias of the standard correlated AIC, which reads, up to model-independent terms, $2n_\eta$.

\section{RIBAIC bias correction}
\label{sec:AppB}
The bias of the Bayesian extension in eq.~\eqref{eq:RIBAIC} can be computed similarly to App.~\ref{sec:AppA}. It reads
\begin{align}
    {\cal B}(\bsy, C,\eta) \equiv \mathbb{E} \Big[2\log {\cal L}(\bsy,C \mid \bstheta_{\rm MAP}^\text{\tiny{$C$}},\eta) + \Delta(\Sigma \mid\bstheta_{\rm MLE}^\text{\tiny{$\Sigma$}},\eta) \Big]  \, .
\end{align}
The second term coincides with eq.~\eqref{eq:bias2}. Under the assumption of normally-distributed data, the first term reads
\begin{align}
\label{eq:RIBAIC_bias2_intermediate}
    \mathbb{E} \left[ 2\log {\cal L}(\bsy,C \mid \bstheta_{\rm MAP}^\text{\tiny{$C$}},\eta) \right] = -\mathbb{E} \left[ \chi_C^2 (\bstheta_{\rm MAP}^\text{\tiny{$C$}}) \right] - n_d\log2\pi -\log\det C \, .
\end{align}
By expanding around the true value $\bstheta_0$, the expectation value of the chi-square in the expression above can be written as
\begin{align}
\label{eq:RIBAIC_chi2_intermediate}
    \mathbb{E} \left[ \chi_C^2 (\bstheta_{\rm MAP}^\text{\tiny{$C$}}) \right] = \tr\left[C^{-1} \Sigma \right] - 2 \mathbb{E} \left[ \delta \bstheta^T J^T(\bstheta_0) C^{-1} \left(\bsy-\eta(\bstheta_0)\right)  \right] + \mathbb{E} \left[ \delta\bstheta^T H_C(\bstheta_0) \delta\bstheta \right] \, ,
\end{align}
where $\delta\bstheta = \bstheta_{\rm MAP}^\text{\tiny{$C$}} - \bstheta_0$ and we used the definition in eq.~\eqref{eq:H}. Note that for the first term on the r.h.s. we used the result for the expectation values of quadratic forms. The remaining two terms are computed analogously, by using the condition at the minimum of the augmented chi-square~\cite{Lepage:2001ym}. The condition at the minimum leads to the relation
\begin{align}
\label{eq:MAP_min}
    \bstheta_{\rm MAP}^\text{\tiny{$C$}} - \bstheta_0 = {\cal H}^{-1}_C(\bstheta_0) J^T(\bstheta_0) C^{-1} \left(\bsy - \eta(\bstheta_0)\right) + {\cal H}^{-1}_C(\bstheta_0) C_\theta^{-1}( \tilde\bstheta - \bstheta_0 ) \, .
\end{align}
Let us stress that the equation above differs from eq.~\eqref{eq:MLE_min} due to the presence of priors and because $\bstheta_{\rm MAP}^\text{\tiny{$C$}}$ does not directly minimize the chi-square of the data appearing in the bias, but rather it minimizes the augmented chi-square. In analogy to eq.~\eqref{eq:H}, in eq.~\eqref{eq:MAP_min} we introduced
\begin{align}
    {\cal H}_C(\bstheta) = H_C(\bstheta) + C_\theta^{-1} \, ,
\end{align}
where $C_\theta$ is the prior covariance matrix.
If $n_\pi<n_\eta$, $C_\theta^{-1}$ is understood to be the $n_\eta\times n_\eta$ matrix obtained
by inverting the $n_\pi\times n_\pi$ prior covariance matrix and setting to zero the rows
and columns associated with those parameters without priors.
By using eq.~\eqref{eq:MAP_min} in the expectation value in eq.~\eqref{eq:RIBAIC_chi2_intermediate}, and by taking advantage of the well-known result for quadratic forms, the bias term in eq.~\eqref{eq:RIBAIC_bias2_intermediate} simplifies to
\begin{align}
    \mathbb{E} &\left[ 2\log {\cal L}(\bsy,C \mid \bstheta_{\rm MAP}^\text{\tiny{$C$}},\eta)  \right] = - n_d \log 2\pi - \log\det C - \tr\left[C^{-1} \Sigma \right] \nonumber \\[.2cm]
    &- \tr\left[C^{-1/2} \Sigma C^{-1/2} {\cal P}_0 \right]  + 2 \tr \left[ C^{-1/2}\Sigma C^{-1/2} {\cal P}_1 \right] - (\tilde\bstheta-\bstheta_0)^T {\cal Q}^{-1} (\tilde\bstheta-\bstheta_0) \, .
\end{align}
Here, similarly to the projector in eq.~\eqref{eq:proj}, we introduced the matrices
\begin{align}
    \label{eq:P0}
    {\cal P}_0 &= C^{-1/2}J \,  {\cal H}_C^{-1} \, H_C \, {\cal H}_C^{-1} \, J^T C^{-1/2}\mid_{\bstheta_0} \,, \\[.2cm]
    \label{eq:P1}
    {\cal P}_1 &= C^{-1/2}J \,  {\cal H}_C^{-1} \, J^T  C^{-1/2}\mid_{\bstheta_0} \,, \\[.2cm]
    {\cal Q}^{-1} & =  C_\theta^{-1} {\cal H}_C^{-1} \, H_C \, {\cal H}_C^{-1} C_\theta^{-1}  \mid_{\bstheta_0} \, .
    \label{eq:Q}
\end{align}
In practice, to compute these terms instead of using the true parameter vector $\bstheta_0$, which is unknown, we use the estimator $\bstheta_{\rm MAP}^\text{\tiny{$C$}}$. This substitution is justified when priors are either wide or centered close to the true values of the parameters.
If priors are misspecified, the MAP estimator is biased and the resulting 
estimate of $\varepsilon(\tilde{\bstheta})$ may be unreliable.
However, the entire derivation relies on the assumption that the model, including priors,
 is correctly specified, which is checked through the quality of the fit before computing the RIBAIC.
Fits with misspecified priors are therefore expected to be discarded based on their quality.
Finally, by combining the expression above with eq.~\eqref{eq:bias2} we obtain the explicit expression of the bias defined
in eq.~\eqref{eq:RIBAIC_bias} and consequently the Information Criterion in eq.~\eqref{eq:RIBAIC2}.
As a final remark, note that by taking the limit of flat priors, both ${\cal P}_0$ and ${\cal P}_1$ reduce to the projector in eq.~\eqref{eq:proj}, while the term involving ${\cal Q}^{-1}$ vanishes and the overall expression for the bias reproduces the one of the RIAIC derived in App.~\ref{sec:AppA}.

\section{Variance of \texorpdfstring{$\Delta_{\text{\tiny{RIAIC}}}$}{Delta RIAIC}}
\label{sec:AppC}
In this appendix we investigate the statistical properties of
\begin{align}
\label{eq:delta_RIAIC}
\Delta_i = {\rm{RIAIC}}_{i} - {\rm{AIC}}_{i} = \delta X_{i} - \mathbb{E} \left[ \delta X_{i} \right]\,,
\end{align}
where we used the shorthand notation $\Delta \equiv \Delta_{\text{\tiny{RIAIC}}}$ and, in contrast to the notation
of the main text, we introduced the label $i$ which identifies the $i$th fitted dataset.
Note that in the last step on the r.h.s. we explicitly used the expressions in eqs.~\eqref{eq:gAIC} and \eqref{eq:AIC}
and introduced, for a fixed fitted dataset, $\delta X_i \equiv \chi^2_{C_i} - \chi^2_{\Sigma_i}$. Here $C_i$ and $\Sigma_i$ denote
the regularized and the fully-correlated covariance matrices of the $i$th dataset respectively.
Similarly to the main text, the expectation value on the r.h.s.
is taken over the true probability distribution of the fitted dataset. From eq.~\eqref{eq:delta_RIAIC}, $\Delta_i$ has vanishing mean; therefore,
given two fitted datasets, the resulting covariance between $\Delta_i$ and $\Delta_j$ is
\begin{align}
\operatorname{Cov} \left[\Delta_i, \Delta_j\right] = \mathbb{E} \left[\Delta_i\Delta_j\right] = \mathbb{E} \left[ \delta X_i \delta X_j \right] - \mathbb{E} \left[ \delta X_i\right] \mathbb{E} \left[ \delta X_j\right] \,,
\end{align}
where in the last step we used the definition in eq.~\eqref{eq:delta_RIAIC} and expanded the expectation values.
As before, we work under the assumptions that the model is correctly specified and that data follow a multivariate Gaussian distribution.
Following the steps of Apps.~\ref{sec:AppA} and \ref{sec:AppB}, by expanding the chi-square variables in $\delta X_i$ and $\delta X_j$
around the true value $\eta(\bstheta_0)$ and recalling that $\mathbb{E} \left[ \bsy\right] = \eta(\bstheta_0)$ for both $i$ and $j$,
standard results for the second and fourth moments of a multivariate normal distribution lead to
\begin{align}
\label{eq:cov_windows}
\operatorname{Cov} \left[\Delta_i, \Delta_j\right] = 2\tr\left[K_i \Sigma_{ij} K_j \Sigma_{ji}\right] \, ,
\end{align}
where $\Sigma_{ij} = \Sigma_{ji}^T$ is an $n_{d_i} \times n_{d_j}$ matrix which contains the correlations between the two fitted
datasets and, by using the notation of App.~\ref{sec:AppA},
\begin{align}
K_i \equiv C_i^{-1/2}\left[1-{\cal P}_i\right]C_i^{-1/2} - \Sigma_i^{-1/2} \left[1-\Sigma^{-1/2}_i J_i H_{\Sigma_i}^{-1} J_i^T \Sigma_i^{-1/2}\right] \Sigma_i^{-1/2} \, ,
\end{align}
where ${\cal{P}}_i$ is the projector in eq.~\eqref{eq:proj} and $J_i$ is the Jacobian, both restricted to the $i$th dataset.
The variance of $\Delta_i$ on the $i$th fitted subset is obtained from the diagonal term $\operatorname{Cov}[\Delta_i,\Delta_i]$ and can be written in the compact form
\begin{align}
\label{eq:variance}
\operatorname{Var} \left[\Delta_i\right] = 2\tr \lb \left[ (1-{\cal P}_i)(C_i^{-1/2}\Sigma_i C_i^{-1/2} - 1)\right]^2 \rb \,.
\end{align}
Note that, as expected, in the correlated limit, i.e. for $C_i=\Sigma_i$, $\Delta_i$ has vanishing variance since it is identically zero.

\section{Software implementation}
A Python implementation of the Information Criteria introduced in this work is publicly
available at \url{https://github.com/laudid46/pyaic} under the GPL-3.0 license.

\printbibliography

@article{Bruno:2022mfy,
    author = "Bruno, Mattia and Sommer, Rainer",
    title = "{On fits to correlated and auto-correlated data}",
    eprint = "2209.14188",
    archivePrefix = "arXiv",
    primaryClass = "hep-lat",
    doi = "10.1016/j.cpc.2022.108643",
    journal = "Comput. Phys. Commun.",
    volume = "285",
    pages = "108643",
    year = "2023"
}

@article{bootstrap,
author = {B. Efron},
title = {{Bootstrap Methods: Another Look at the Jackknife}},
volume = {7},
journal = {The Annals of Statistics},
number = {1},
publisher = {Institute of Mathematical Statistics},
pages = {1 -- 26},
year = {1979},
doi = {10.1214/aos/1176344552},
URL = {https://doi.org/10.1214/aos/1176344552}
}

@article{Borsanyi:2020mff,
    author = "Borsanyi, Sz. and others",
    title = "{Leading hadronic contribution to the muon magnetic moment from lattice QCD}",
    eprint = "2002.12347",
    archivePrefix = "arXiv",
    primaryClass = "hep-lat",
    doi = "10.1038/s41586-021-03418-1",
    journal = "Nature",
    volume = "593",
    number = "7857",
    pages = "51--55",
    year = "2021"
}

@article{10.1214/aoms/1177729694,
author = {S. Kullback and R. A. Leibler},
title = {{On Information and Sufficiency}},
volume = {22},
journal = {The Annals of Mathematical Statistics},
number = {1},
publisher = {Institute of Mathematical Statistics},
pages = {79 -- 86},
year = {1951},
doi = {10.1214/aoms/1177729694},
URL = {https://doi.org/10.1214/aoms/1177729694}
}

@article{https://doi.org/10.1002/wics.1460,
author = {Cavanaugh, Joseph E. and Neath, Andrew A.},
title = {The Akaike information criterion: Background, derivation, properties, application, interpretation, and refinements},
journal = {WIREs Computational Statistics},
volume = {11},
number = {3},
pages = {e1460},
doi = {10.1002/wics.1460},
year = {2019}
}

@article{Jay:2020jkz,
    author = "Jay, William I. and Neil, Ethan T.",
    title = "{Bayesian model averaging for analysis of lattice field theory results}",
    eprint = "2008.01069",
    archivePrefix = "arXiv",
    primaryClass = "stat.ME",
    reportNumber = "FERMILAB-PUB-20-374-T",
    doi = "10.1103/PhysRevD.103.114502",
    journal = "Phys. Rev. D",
    volume = "103",
    pages = "114502",
    year = "2021"
}

@article{Akaike1,
    author = "Akaike, H.",
    title = "{A new look at the statistical model identification}",
    doi = "10.1109/TAC.1974.1100705",
    journal = "IEEE Trans. Automatic Control",
    volume = "19",
    number = "6",
    pages = "716--723",
    year = "1974"
}

@Inbook{Akaike2,
author="Akaike, Hirotugu",
editor="Parzen, Emanuel
and Tanabe, Kunio
and Kitagawa, Genshiro",
title="Information Theory and an Extension of the Maximum Likelihood Principle",
bookTitle="Selected Papers of Hirotugu Akaike",
year="1998",
publisher="Springer New York",
address="New York, NY",
pages="199--213",
isbn="978-1-4612-1694-0",
doi="10.1007/978-1-4612-1694-0_15",
url="https://doi.org/10.1007/978-1-4612-1694-0_15"
}

@misc{Frison:2023lwb,
      title={Towards fully bayesian analyses in Lattice QCD}, 
      author={Julien Frison},
      year={2023},
      eprint={2302.06550},
      archivePrefix={arXiv},
      primaryClass={hep-lat},
      url={https://arxiv.org/abs/2302.06550}, 
}

@article{Neil:2022joj,
    author = "Neil, Ethan T. and Sitison, Jacob W.",
    title = "{Improved information criteria for Bayesian model averaging in lattice field theory}",
    eprint = "2208.14983",
    archivePrefix = "arXiv",
    primaryClass = "stat.ME",
    doi = "10.1103/PhysRevD.109.014510",
    journal = "Phys. Rev. D",
    volume = "109",
    number = "1",
    pages = "014510",
    year = "2024"
}

@article{Neil:2023pgt,
    author = "Neil, Ethan T. and Sitison, Jacob W.",
    title = "{Model averaging approaches to data subset selection}",
    eprint = "2305.19417",
    archivePrefix = "arXiv",
    primaryClass = "stat.ME",
    doi = "10.1103/PhysRevE.108.045308",
    journal = "Phys. Rev. E",
    volume = "108",
    number = "4",
    pages = "045308",
    year = "2023"
}

@article{Lepage:2001ym,
    author = "Lepage, G. P. and Clark, B. and Davies, C. T. H. and Hornbostel, K. and Mackenzie, P. B. and Morningstar, C. and Trottier, H.",
    editor = "Muller-Preussker, M. and Bietenholz, Wolfgang and Jansen, K. and Jegerlehner, F. and Montvay, I. and Schierholz, G. and Sommer, R. and Wolff, U.",
    collaboration = "HPQCD",
    title = "{Constrained curve fitting}",
    eprint = "hep-lat/0110175",
    archivePrefix = "arXiv",
    reportNumber = "FERMILAB-CONF-01-485-T",
    doi = "10.1016/S0920-5632(01)01638-3",
    journal = "Nucl. Phys. B Proc. Suppl.",
    volume = "106",
    pages = "12--20",
    year = "2002"
}

@article{Wolff:2003sm,
    author = "Wolff, Ulli",
    collaboration = "ALPHA",
    title = "{Monte Carlo errors with less errors}",
    eprint = "hep-lat/0306017",
    archivePrefix = "arXiv",
    reportNumber = "HU-EP-03-32, SFB-CPP-03-12",
    doi = "10.1016/S0010-4655(03)00467-3",
    journal = "Comput. Phys. Commun.",
    volume = "156",
    pages = "143--153",
    year = "2004",
    note = "[Erratum: Comput.Phys.Commun. 176, 383 (2007)]"
}

@inproceedings{BaSc:2026fdy,
    author = "Raposo, Andr{\'e} Bai{\~a}o and others",
    collaboration = "BaSc",
    title = "{Preliminary study of the $H$ dibaryon in $N_{\rm f}=2+1$ lattice QCD}",
    booktitle = "{42nd International Symposium on Lattice Field Theory}",
    eprint = "2603.00698",
    archivePrefix = "arXiv",
    primaryClass = "hep-lat",
    reportNumber = "DESY-26-028",
    month = "2",
    year = "2026"
}

@article{BaryonScattering:2026voj,
    author = "Baeza-Ballesteros, Jorge and others",
    collaboration = "Baryon Scattering",
    title = "{Two-nucleon scattering amplitudes at light quark masses from lattice QCD}",
    eprint = "2610.23332",
    archivePrefix = "arXiv",
    primaryClass = "hep-lat",
    reportNumber = "DESY-26-136",
    month = "10",
    year = "2026"
}

@article{Kuberski:2024pms,
    author = {Kuberski, Simon and Joswig, Fabian and Collins, Sara and Heitger, Jochen and S{\"o}ldner, Wolfgang},
    collaboration = "RQCD, ALPHA",
    title = "{D and D$_{s}$ decay constants in N$_{f}$ = 2 + 1 QCD with Wilson fermions}",
    eprint = "2405.04506",
    archivePrefix = "arXiv",
    primaryClass = "hep-lat",
    reportNumber = "CERN-TH-2024-052, MITP-24-047, MS-TP-24-09",
    doi = "10.1007/JHEP07(2024)090",
    journal = "JHEP",
    volume = "07",
    pages = "090",
    year = "2024"
}

@software{gvar,
      author       = {Peter Lepage},
      title        = {gplepage/gvar},
      publisher    = {GitHub},
      month        = apr,
      year         = 2020,
      note         = {\url{https://github.com/gplepage/gvar}}
}

@software{lsqfit,
     author       = {Peter Lepage},
     title        = {gplepage/lsqfit},
     publisher    = {GitHub},
     month        = apr,
     year         = 2020,
     note         = {\url{https://github.com/gplepage/lsqfit}}
}

@article{Baeza-Ballesteros:2026qxo,
    author = "Baeza-Ballesteros, Jorge and others",
    title = "{Quark-mass dependence of two-nucleon scattering}",
    eprint = "2611.07864",
    archivePrefix = "arXiv",
    primaryClass = "hep-lat",
    reportNumber = "DESY-26-150",
    month = "11",
    year = "2026"
}
\end{document}